\documentclass[trackchange,twocolumn]{aastex701}
\usepackage{amsmath} 

\begin{document}

\title{CO and N$_{2}$ Produced from H$_{2}$O, CO$_{2}$, and NH$_{3}$ Cometary Ice Analogs}

\author[0000-0002-2295-5452]{Alexandra McKinnon}
\affiliation{Department of Chemistry and Chemical Biology, Harvard University, 12 Oxford Street, Cambridge, MA 02138, USA}
\email{alexandramckinnon@g.harvard.edu}

\author[0000-0002-5061-3054]{Alexia Simon}
\affiliation{Center for Astrophysics $|$ Harvard \& Smithsonian, 60 Garden Street, Cambridge, MA 02138, USA}
\email{alexia.simon@cfa.harvard.edu}

\author[0000-0002-9889-3921]{Michelle R. Brann}
\affiliation{Center for Astrophysics $|$ Harvard \& Smithsonian, 60 Garden Street, Cambridge, MA 02138, USA}
\email{michelle.brann@cfa.harvard.edu}

\author[0000-0001-6947-7411]{Elettra L. Piacentino}
\affiliation{Center for Astrophysics $|$ Harvard \& Smithsonian, 60 Garden Street, Cambridge, MA 02138, USA}
\email{elettra.piacentino@cfa.harvard.edu}

\author[0000-0001-8798-1347]{Karin I. Öberg}
\affiliation{Center for Astrophysics $|$ Harvard \& Smithsonian, 60 Garden Street, Cambridge, MA 02138, USA}
\email{koberg@cfa.harvard.edu}

\author[0000-0003-2761-4312]{Mahesh Rajappan}
\affiliation{Center for Astrophysics $|$ Harvard \& Smithsonian, 60 Garden Street, Cambridge, MA 02138, USA}
\email{mrajappan@cfa.harvard.edu}

\begin{abstract}
Hypervolatile species such as carbon monoxide (CO) and molecular nitrogen (N$_{2}$) have been detected in comets, and could be used to constrain comet formation temperature conditions if their presence is due to freeze-out and/or entrapment. Here we instead explore another plausible origin of cometary hypervolatiles: photodissociation of less volatile species. We characterize CO and N$_{2}$ formation following ultraviolet (UV) irradiation and electron bombardment of carbon dioxide (CO$_{2}$), ammonia (NH$_{3}$), H$_{2}$O:CO$_{2}$, H$_{2}$O:NH$_{3}$, and H$_{2}$O:CO$_{2}$:NH$_{3}$ cometary ice analogs. We find that CO and N$_{2}$ form in all photoprocessed ices at temperatures between 10~K and 100~K, resulting in  0.4-0.9~$\%$ CO and 0.03-0.7~$\%$ N$_2$ relative to water, and CO/CO$_2$ and N$_2$/NH$_3$ mixing ratios of 2.5-62~$\%$ and 0.7-9~$\%$, respectively, across the experiments. Because our initial ices are reasonably well-matched to interstellar ices and we use UV exposure similar to a dark cloud, we can compare the resulting ratios directly to cometary abundances. Such a comparison shows that while only a few of CO observations in comets are readily explained by photodissociation, almost all observed cometary N$_2$ can be accounted for by photodissociation of NH$_3$ embedded in water ice. The latter result is also consistent with observed similarly elevated isotopic ratios of N$_{2}$ and NH$_{3}$ in 67P. Taken together, our results suggest that N$_2$/H$_2$O ratios $<$1~\% should be used cautiously when inferring a comet's formation location, while the more substantial CO abundances seen in many comets do likely imply entrapment at low ice temperatures. 
\end{abstract}


\section{Introduction} \label{sec:intro}
The distribution of volatiles in planet-forming disks ultimately determines the compositions of planets. Our own planet-forming disk is largely gone today, but the composition of the outer disk can still be inferred from comet abundances, since most models and observations suggest that comets largely preserve the chemical compositions present at their formation \citep{mumma2011chemical,weissman2020origin,altwegg2019cometary}. A combination of remote observations and missions have determined that comets often contain hypervolatiles like carbon monoxide (CO) and molecular nitrogen (N$_{2}$) \citep{iro2003interpretation,cochran2015composition,le2015inventory,rubin2015molecular,rubin2019elemental,mckay2019peculiar}. Hypervolatiles only freeze out from the gas-phase at low temperatures (see below), and their presence therefore indicates that the comet building blocks (icy dust grains) assembled at correspondingly low temperatures, entrapping the hypervolatiles in the icy grain mantles \citep{rubin2015molecular}. 
Entrapment is the phenomenon in which volatile molecules are trapped in less volatile molecules, such as amorphous water. When entrapped, highly volatile molecules are prevented from desorbing off of the ice films \citep{collings2004laboratory}. The entrapment of hypervolatiles from the gas-phase has been found to be very efficient at low temperatures, \citep{bar1985trapping,bar1988trapping,notesco1996enrichment,collings2004laboratory,simon2023entrapment,gudipati2023thermal,kipfer2024sublimation} but drops off rapidly above 30~K \citep{bar2007trapping,zhou2024competitive}. 

Entrapment of hypervolatiles formed in the gas-phase is not the only possible explanation for the presence of hypervolatiles in comets, however. Previous studies have proposed that substantial amounts of cometary O$_2$ could originate from photodissociation chemistry inside of icy grains, using less volatile ice constituents as precursors \citep{bulak2022quantification}. While the plausibility of this O$_2$ formation mechanism is under debate \citep{mousis2016protosolar,bieler2015abundant,taquet2016primordial,altwegg2019cometary}, it suggests that we should consider whether (some)  cometary CO and N$_2$ could be explained by the decomposition of less volatile species in the icy grain precursors to comets, relaxing constraints on the formation conditions of CO and N$_{2}$-containing comet material. Such photolysis would have to take place in small enough grains that their icy mantles are transparent to UV, $\sim\mu$m-sized grains, and hence be limited to the earliest stages of planetesimal formation in protoplanetary disks. We further note that for the produced CO and N$_2$ to remain in the ice during subsequent collisional growth and pebble drift into warmer disk regions will require a kind of entrapment; the complete question is then whether (1) sufficient hypervolatiles can plausibly form through photolysis, and (2) whether hypervolatiles formed from irradiation can become `entrapped' into icy grain mantles at higher temperatures, compared to the entrapment of hypervolatiles from the gas-phase.

With the exception of CO-dominated comets such as C/2016 R2 and 2I/Borisov (\citep{mckay2019peculiar} and references therein), the CO  abundances in comets with respect to water range between $<$1~$\%$ and 32~$\%$ \citep{le2015inventory,rubin2015molecular}. The most well-studied comet, Comet 67P, falls towards the lower portion of the CO abundance range, having an abundance of 3~$\%$ with respect to water \citep{lauter2020gas}. Before the Rosetta mission to Comet 67P, N$_{2}$ abundances with respect to water had only been reported a handful of times in comets, based on N$_{2}$$^{+}$ optical measurements \citep{cochran2000n+,cochran2002search,cochran2015composition}, and ranged between $<$0.0016~$\%$ and 0.7~$\%$ with respect to water \citep{iro2003interpretation,anderson2023n2}. Comet 67P has an N$_{2}$ bulk abundance of 0.089~$\pm$~0.024~$\%$ \citep{rubin2019elemental}, consistent with N$_{2}$$^{+}$ measurements towards other comets. As for many other comets, the substantial abundances of CO and N$_{2}$ in the coma of 67P have been used to infer a cold origin of the 67P building blocks, assuming that the CO and N$_{2}$ were entrapped in water ice \citep{rubin2015molecular}. 

However, as mentioned above, hypervolatile comet constituents may also be produced through energetic processes in interstellar and/or disk ice grains \citep{chen2011photo}. Icy grains are first formed in the interstellar medium, before the onset of star formation. These grains are observed to be rich in CO$_2$ and NH$_3$, which may act as precursors of CO and N$_2$. 

Based on observations with Spitzer and JWST of ice compositions in clouds, and protostellar envelopes, and disks, \citep{boogert2015observations,mcclure2023ice} CO$_2$ and NH$_3$ are present at abundances relative to water of 12\%-50\% and 3\%-10\%, respectively. The average low-mass protostellar H$_2$O:CO$_2$:NH$_3$ mixing ratio is 100:28:6 \citep{boogert2015observations}, with all NH$_3$ and $\sim$2/3 of the CO$_2$ likely residing together with water \citep{pontoppidan2008c2d}. Both CO$_2$ and NH$_3$ have also recently been detected in a protoplanetary disk, albeit at somewhat lower abundances of 15\% for CO$_2$ and 1.7\% for NH$_3$, respectively \citep{mcclure2023ice,sturm2024jwst}. When embedded in water ice, these mid-volatility species are likely to survive disk formation when only considering thermal effects \citep{visser2009photodissociation,drozdovskaya2016cometary}. CO$_2$ and NH$_3$ are also detected in comets, but at lower abundances than would be expected from unperturbed inheritance; in comet 67P, the abundances of CO$_{2}$ and NH$_{3}$ with respect to water are 7~$\%$ and 0.4~$\%$, respectively \citep{lauter2020gas}. While we cannot rule out that these lower abundances reflect unusual starting conditions for the Solar Nebula and/or an unusually hot accretion phase, the assumption for the remainder of this paper, unless otherwise noted, is that these comet abundances are low due to photolysis of icy grain mantles in the protostellar or protoplanetary disk phase; based on published photodissociation cross sections both CO$_2$ and NH$_3$ should be consumed when exposed to $\sim$ 10$^{18}$ photons cm$^{-2}$, which is reached in 3~Myrs in a molecular cloud or 3,000 years in the upper molecular layers of a typical protoplanetary disk, assuming a disk UV flux of 10$^{8}$~cm$^{-2}$~s$^{-1}$ \citep{Bergner2021}. Refining our question further, we can then ask how much CO and N$_2$ can be produced through photodissociation of CO$_2$ and NH$_3$ ice and subsequently retained through entrapment, given the assumption that Solar System ices began as typical protostellar ones.

Laboratory ice experiments have been explored to investigate reaction pathways on grain surfaces for decades \citep{d1986time,sandford1988laboratory,sandford1990volume,schutte1993experimental,bernstein1995organic,allamandola1999evolution}. Specifically, \citet{allamandola1988photochemical} pioneered irradiation experiments of cometary ice analogs. Since then, previous experiments show that qualitatively CO and N$_2$ are produced from CO$_2$ and NH$_3$ in interstellar ice analogs under a range of ice conditions. \citet{martin2015uv}, \citet{bulak2022quantification}, \citet{potapov2022formation}, and \citet{chen2011photo} reported the formation of CO when pure CO$_{2}$ ice (8~K), H$_{2}$O:CO$_{2}$ ice mixtures (20~K), 5:4:1 H$_{2}$O:NH$_{3}$:CO$_{2}$ ices (75--150~K), and 1:1:1 H$_{2}$O:CO$_{2}$:NH$_{3}$ ices (16~K) are irradiated. In these experiments up to 35~$\%$ of an undiluted CO$_{2}$ ice can be converted into CO \citep{martin2015uv} but there was no quantitative CO yield reported from water-rich binary and ternary ices irradiation. \citet{martin2018uv}, \citet{loeffler2010radiation}, \citet{zheng2010formation}, and \citet{chen2011photo} also reported the production of N$_{2}$ from UV irradiation of undiluted NH$_{3}$ ices, from proton bombardment of a 2:1 H$_{2}$O:NH$_{3}$ ice, from UV irradiation a 10:1 H$_{2}$O:NH$_{3}$ ice, and from UV irradiation of 1:1:1 H$_{2}$O:CO$_{2}$:NH$_{3}$ ices, but did not report a yield. As a collection, these experiments show that some CO and N$_2$ are indeed produced when interstellar ice analogs are exposed to UV or electron irradiation, but it is currently unclear how much of a comet's hypervolatile reservoir could be formed through such ice photolysis. At present, we hence do not know under which conditions a cometary CO and N$_2$ reservoir can be explained by icy grain chemistry, and when it must be attributed to entrapment of gas-phase molecules or hypervolatile freeze-out, and therefore can be used to say something about the comet formation temperature. 

\indent  In this paper we explore how the CO and N$_{2}$ formation from CO$_2$ and NH$_3$ ice depends on the ice composition and temperature, using protostellar ice analogs as our starting point. The experimental methods used to measure photolytic CO and N$_2$ production are described in Section~\ref{sec:methods}. The results are presented and discussed in Section~\ref{sec:results}. Section~\ref{sec:astrophys} presents the implications of the results on our interpretation of cometary abundances. The findings are summarized in Section~\ref{sec:conclusions}.

\section{Methods} \label{sec:methods}
\subsection{Experimental Set-up}\label{subsec:experimental}

The experiments were performed on the Surface Processing Apparatus for Chemical Experimentation to Constrain Astrophysical Theories (SPACECAT), which has been previously described in detail \citep{lauck2015co,bergner2017methanol,martin2020formation}. Briefly, the ices are prepared in an ultra-high vacuum (UHV) chamber with a base pressure of approximately 1~$\times 10^{-10}$ Torr on a suspended CsI substrate, which is cooled to the temperature of interest by a closed-cycle helium cryocooler (DE204B, Advanced Research Systems, Inc.). The temperature is monitored by a Lakeshore 335 temperature controller, which has an accuracy within 2~K. The chamber is further equipped with a Bruker Vertex~70v Infrared (IR) Spectrometer, as well as a Pfeiffer QMG 220M1 Quadrupole Mass Spectrometer (QMS, 0.5~amu resolution), which are used to monitor the ice and gas compositions, respectively. We irradiate the ice with an H$_{2}$D$_{2}$ lamp (Hamamatsu, L11798). The lamp intensity peaks at $\sim$161~nm and has a measured photon flux of about 1~$\times 10^{14}$~cm$^{-2}$~s$^{-1}$ and we measure this with a AXUV-100G photodiode calibrated by NIST.
%\citep{bergner2017methanol}.
\newline \indent A subset of our experiments were also run on the Surface Photoprocessing Apparatus Creating Experiments To Investigate
Grain Energetic Reactions (SPACETIGER) to explore  CO and N$_2$ production when the ice is instead exposed to electron bombardment. This experimental setup has also been previously described in \citet{maksyutenko2022formation}. SPACETIGER utilizes a different model of cryocooler (DE210B), and temperature controller (Lakeshore model 336). SPACETIGER is also equipped with a copper substrate which is required to ground the sample upon 2~keV electron bombardment. As copper is opaque, the IR spectrometer on SPACETIGER is used in reflection-absorption mode, and we used a reflection angle of 60$^{\circ}$. The ice mixtures were electron irradiated using an ELG-2/EGP5-1022 electron source from Kimball Physics. The electron energy was set to 2~keV, the electron beam current was approximately 0.7~$\mu$A, and the typical irradiation time was 120~minutes.

\subsection{Experimental Procedures}\label{sec:Gas}

We first prepare the gas sample in the gas line at ambient temperature. The molecules of interest (H$_{2}$O: deionized; $^{13}$CO$_{2}$: 99\%, Sigma-Aldrich; $^{13}$CO: 99\%, Sigma-Aldrich; $^{15}$NH$_{3}$: 98\%, Sigma-Aldrich) were all prepared as pure (undiluted) or mixed gas samples prior to each experiment. We use the $^{13}$CO$_2$ and $^{15}$NH$_{3}$ isotopologues to avoid confusion with any residual $^{12}$CO and/or $^{14}$N$_{2}$ in the chamber, both of which would appear at m/z=28 on the QMS. We confirm that our isotopologue choice for CO$_2$ has no major impact in Appendix~\ref{app:12CO} and hereafter refer to $^{13}$CO$_2$ and $^{13}$CO as CO$_2$ and CO, respectively for increased readability.  For binary and ternary mixtures, we prepare appropriate volumes of each gas in an isolated cylinder and then mix the separated components prior to dosing for about 10~minutes. We then dose the gas mixture onto a cryocooled CsI substrate through a variable leak valve at 10~K and at a dosing rate of about 17~monolayers per minute. We then use FTIR spectroscopy at 10~K (in transmission mode for our UV experiments and in reflection mode for our electron bombardment experiments) to characterize the ice composition. We determine the column densities by integrating the relevant spectral band, and applying literature band strengths from Table~\ref{tbl:BandStrengths} in Appendix~\ref{app:IR_Spectra}. We report ice column densities in units of monolayers (ML), under the assumption that 1~ML~= 10$^{15}$ molecules~cm$^{-2}$.

We next expose the ice to UV light until we reach or at least approach steady state, which in the majority of  our experiments requires $\sim$~1~$\times~10^{18}$ FUV photons cm$^{-2}$. %If we approximate each photon from the H$_2$D$_2$ lamp has about 8~eV, this results in a total dose of $\sim10^{19}$~eV, which 
A similar fluence is reached after $\sim$3~Myrs in a dark cloud core assuming a background UV field of $\sim10^4$ cm$^{-2}$ s$^{-1}$ \citep{shen2004cosmic}. Of more direct relevance to the questions posed by this paper $\sim$~1~$\times~10^{18}$ FUV photons cm$^{-2}$ is considerably lower than the UV fluence typical small grains will experience exterior to the water ice line. The upper layers of disks are irradiated with relatively high fluxes (the relevant UV fluence is achieved at intermediate disk heights in $<$1,000 yrs), and while the bulk of icy grains reside further down in the midplane most of the time, they are readily lofted into the molecular layers as long as there is some turbulent mixing \citep{Bergner2021}, resulting in considerable UV exposure, especially for smaller grains beyond the water ice line.

 For the electron bombardment experiments, we exposed the ice to an electron dose beyond what was sufficient to reach steady state, and achieved typical total fluences of $\sim$4~$\times~10^{16}$ electrons cm$^{-2}$. The area of our substrate is 0.78~cm$^{2}$; thus this corresponds to an energy dose of $\sim$3~$\times~10^{16}$ electrons, or $\sim$~6~$\times~10^{19}$ eV considering that we use 2~keV electrons. This can be compared with the FUV doses of $\sim$~8~$\times~10^{18}$~eV achieved in the FUV experiments considering a typical UV photon energy of 8~eV and our diode used to measure the photon current has an area of 1~cm$^{2}$. The UV spectrum of our lamp is presented in \citet{piacentino2025survival}. Throughout irradiation, IR scans are collected regularly to monitor any changes in the spectral features of the ice. 

Following irradiation, the ice is slowly warmed with Temperature Programmed Desorption (TPD) to 200~K at a rate of 2~K per minute. During the TPD experiment, we use the QMS to identify the mass-to-charge ratio of any desorbed species, which is key to measure the amount of formed N$_2$. The experiments are listed in Table~\ref{tbl:experiments}.

\begin{deluxetable}{lccc}[htbp]										
\tabletypesize{\scriptsize}											
\tablewidth{0pt}													
%\tablenum{1}													
\tablecaption{The experiments performed in this study. 
\label{tbl:experiments}}											
\tablehead{\colhead{Molecules} & \colhead{N$_0$*} & \colhead{Irr.} & \colhead{Fluence$^{\star}$}\\							
\colhead{} & \colhead{(ML)} & \colhead{Temp (K)} & \colhead{(10$^{18}$ cm$^{-2}$)}}																							
\startdata													
\hline %\\													
SPACECAT \\													
\hline% \\								
$^{13}$CO	&	23	&	-	&	-	\\						
$^{13}$CO$_{2}$	&	31	&	-	&	-	\\						
$^{13}$CO$_{2}$	&	22	&	10	&	1.17	\\			%		2	\\
\hline	%\\						\\						
$^{15}$NH$_{3}$	&	14	&	-	&	-	\\						
$^{15}$NH$_{3}$	&	7	&	10	&	1.17	\\							\hline	%\\						\\						
H$_{2}$O:$^{13}$CO$_{2}$	&	97:14	&	-	&	-	\\						
H$_{2}$O:$^{13}$CO$_{2}$	&	105:17	&	10	&	1.17	\\%					0.1	\\
\hline	%\\		%				\\						
H$_{2}$O:$^{15}$NH$_{3}$	&	104:13	&	-	&	-	\\						
H$_{2}$O:$^{15}$NH$_{3}$	&	102:9	&	10	&	1.17	\\						
\hline	%\\	%					\\						
H$_{2}$O:$^{13}$CO$_{2}$:$^{15}$NH$_{3}$	&	110:17:6	&	-	&	-	\\						
H$_{2}$O:$^{13}$CO$_{2}$:$^{15}$NH$_{3}$$^b$	&	119:17:4	&	10	&	1.14	\\		%			0.3	\\
H$_{2}$O:$^{13}$CO$_{2}$:$^{15}$NH$_{3}$$^b$	&	96:20:5	&	10	&	1.31	\\	%				0.2	\\
H$_{2}$O:$^{13}$CO$_{2}$:$^{15}$NH$_{3}$$^a$	&	97:15:5	&	10	&	1.26	\\	%				0.2	\\
H$_{2}$O:$^{13}$CO$_{2}$:$^{15}$NH$_{3}$	&	122:16:5	&	30	&	1.18	\\	%				0.1	\\%2025/04/08
H$_{2}$O:$^{13}$CO$_{2}$:$^{15}$NH$_{3}$	&	115:12:4	&	50	&	1.13	\\	%				0.2	\\
H$_{2}$O:$^{13}$CO$_{2}$:$^{15}$NH$_{3}$	&	123:13:5	&	100	&	1.02	\\%					0.2	\\
\hline					%\\						
\hline				%\\						
SPACE	TIGER	&		&		&	\\						
\hline%	\\					%	\\						
$^{13}$CO$_{2}$	&	17	&	10	&	0.0360	\\	%				2	\\
\hline	%\\					%	\\						
$^{15}$NH$_{3}$	&	47	&	10	&	0.0360	\\						
\hline	%\\					%	\\						
H$_{2}$O:$^{13}$CO$_{2}$	&	64:9	&	10	&	0.0344	\\					%0.1	\\
\hline	%\\				%		\\						
H$_{2}$O:$^{15}$NH$_{3}$	&	100:9	&	10	&	0.0367	\\						
\hline	%\\				%		\\						
H$_{2}$O:$^{13}$CO$_{2}$:$^{15}$NH$_{3}$	&	113:12:5	&	10	&	0.0355	\\					%0.2	\\
H$_{2}$O:$^{13}$CO:$^{15}$NH$_{3}$	&	56:5:5	&	-	&	-	\\
\enddata													
\tablecomments{*An error of 20~$\%$ is used for the band strengths of each ice component for N$_0$. $^a$Fiducial Experiment; $^b$Repeat Experiment. $^{\star}$Fluence for SPACECAT experiments is in units of photons~$\times$~cm$^{-2}$ and fluence for SPACETIGER experiments is in units of electrons~$\times$~cm$^{-2}$.}
\end{deluxetable}	

\subsection{Infrared Spectra Analysis} \label{sec:IR}

\begin{deluxetable}{cccc}[htbp]
\tabletypesize{\scriptsize} 
\tablewidth{0pt} 
%\tablenum{1}
\tablecaption{Summary of the peaks used for integration as well as their band strengths. \label{tbl:BandStrengths}}
\tablehead{
\colhead{Substance} & \colhead{Transition Used} & \colhead{Wavenumber} &\colhead{Column Density}\\
&& \colhead{(cm$^{-1}$)} &\colhead{(cm$\times$molecule$^{-1}$})}
\startdata
H$_{2}$O&OH Stretch&3280&$2.2\times10^{-16}$\\
%H$_{2}$$^{18}$O&O-H Stretch&3280&$2.2\times10^{-16}$\\
$^{13}$CO$_{2}$&C=O Stretch&2283&$1.15\times10^{-16}$\\
$^{13}$CO &C-O Stretch&2092&$1.7\times10^{-17}$\\
NH$_{3}$&NH$_{3}$ Umbrella&1070&$1.95\times10^{-17}$\\
\enddata
\tablecomments{H$_{2}$O, $^{13}$CO$_{2}$, and $^{13}$CO band strength from \citet{gerakines1994infrared}, with density corrections from \citet{bouilloud2015bibliographic}. For the $^{15}$NH$_{3}$ band strength we assume minimal change from the $^{14}$NH$_{3}$ band strength reported by \citet{hudson2022ammonia}. For every band strength, we assume an uncertainty of 20\%.}
\end{deluxetable}

All IR spectra obtained as a part of this project are available on Zenodo [10.5281/zenodo.17704002 and 10.5281/zenodo.17704716]. For IR active species (CO$_2$, CO, NH$_3$) we calculate column densities (N$_i$) as:

\begin{equation}\label{eqn:monolayers}
N_i = \frac{\zeta}{A} \int_{band} \tau_{\nu}~d\nu
\end{equation}

$A$ represents the band strength of the transition and $\tau_{\nu}$ is the spectral feature optical depth, and are presented in Table~\ref{tbl:BandStrengths}.
$\zeta$ represents the correction factor required for viewing IR spectra in reflection-absorption mode as opposed to in transmission mode, where $\zeta$~=~1. For the reflection-absorption experiments we can estimate $\zeta$  based on the reflection angle of 60$^\circ$, which yields $\zeta\sim 0.5$. We tested this experimentally by dosing CO on the Cu substrate and repeating this on the CsI substrate under identical conditions and obtained an experimental $\zeta$~=~0.53. 

We integrate the CO$_{2}$ 2283~cm$^{-1}$ peak numerically after subtracting a linear baseline. We fit  both the CO 2092~cm$^{-1}$ band and the NH$_{3}$ 1070~cm$^{-1}$ band with a Gaussian function, and integrate to obtain the integrated optical depth  after subtracting a linear baseline. We estimate the  column density measurement uncertainties based on uncertainties in \textit{A} values, errors on baseline fits, and errors on curve integrations, but in reality the \textit{A} value uncertainties of $\sim$20\% \citep{bouilloud2015bibliographic} always dominate. These estimated uncertainties include possible changes in band strengths due to phase changes following exposure to radiation \citep{schrauwen2025infrared}, and changes in ice matrix. For our ices reported differences in band strengths due to phase changes are up to 10\%: \citet{hudson2022ammonia} found a 10\% difference in band strengths of the relevant NH$_3$ band in  crystalline and amorphous ice, \citet{hudson2025carbon} report a 9\% difference between amorphous $^{12}$CO$_2$ and $^{12}$CO band strengths, and \citet{bouilloud2015bibliographic} did not find a measurable difference between amorphous and crystalline $^{12}$CO ice. The impact of the matrix on the relevant band strengths can be larger, up to 20\%: \citet{bouilloud2015bibliographic} concluded that the band strength of ammonia changes by $\sim$4\% between a water ice matrix compared to its pure, undiluted state, while \citet{mate2025infrared} recently reported that the CO and CO$_2$ band strengths can change by $\sim$20\% between various water-rich mixtures, though the impact on the stretching modes for CO and CO$_2$, which we use, may be smaller.

Though small by comparison, we add a spectral fit uncertainty (which is typically no larger than 4\%) to the band strength uncertainties in quadrature:

\begin{equation}
   \sigma_{{N\mathrm{_{IR}}(x)}} = \sqrt{(\sigma_{Spectralfit})^2+(\sigma_{A})^2}
\end{equation}
which results in typical final column density uncertainties of $\sim$21\% for  CO, $\sim$20\% for CO$_2$, and 20\% for NH$_3$. Throughout the paper we use the final measured NH$_3$, CO$_2$ and CO column densities in each irradiation experiment, together with the N$_2$ column density extracted from mass spectrometric data (see next subsection), to calculate the CO and N$_2$ formation efficiencies.

We model the destruction and formation time series of IR-active ice species using exponential decay and growth curves, respectively. These models are primarily used to assess how close we are to steady-state by the end of each experiment, and in general we continue irradiation until we reach within 10\% of the steady state value. This ensures that the reported CO and N$_2$ formation efficiencies are no longer dependent on the details of the irradiation fluence that the ice has been exposed to, as long as it is similar or higher than the fluence reported for the experiment. Following \citet{jones2011mechanistical}, we model the loss of CO$_{2}$ and NH$_{3}$ as:

\begin{equation}\label{eqn:decay}
N_{(x)}(\Phi) = N_{0(x)}e^{-\Phi\sigma_d}+N_{ss}
\end{equation}
in which $N_{ss}$ is the steady-state column density, $\Phi$ is the fluence in cm$^{-2}$, N$_{0}$ is the initial column density in units of ML, and $\sigma_d$ is the destruction cross section. Our model fit parameters are listed in Table~\ref{tbl:fits} in Appendix~\ref{app:FitParameters}. Beyond comparing measured column densities and modeled steady state column densities, we also we use these fits to assess ice destruction cross sections and how they compare with literature values.

Following \citet{bergner2019oxygen}, the production of CO is modeled as

\begin{equation}\label{eqn:production}
N_{\mathrm{^{13}CO}}~(\Phi) = N_{ss}(1-e^{-\Phi\sigma_f})
\end{equation}
where $\sigma_{f}$ is the formation cross section and $N_{ss}$ is the CO steady state column density, which is always similar to the final CO column density measured at the end of the experiment.  

\subsection{TPD Analysis} \label{sec:QMS}

For species that are IR-inactive, in our case N$_{2}$, we can use calibrated TPD curves to obtain the column density at the end of the experiment. All acquired TPD curves are available in full on Zenodo [10.5281/zenodo.17704834]. As m/z~=~30 exclusively corresponds to $^{15}$N$_{2}$ in these experiments, the number of molecules desorbing from the substrate is linearly proportional to integrated signal from the m/z~=~30 TPD curve. (For readability we will hereafter refer to $^{15}$N$_{2}$ as N$_{2}$) Following \citet{martin2015uv} and \citet{simon2023entrapment}, we calibrated these curves using a blank, unirradiated CO TPD (Figure~\ref{fgr:TPDsBlank} in Appendix~\ref{app:TPDs}), since CO can be measured with both the IR and the QMS spectrometers. For our CO calibration TPD, we fit our data with a linear baseline and then we integrate the area under the m/z=29 ($^{13}$CO) curve numerically in the range of 15~K to 185~K to obtain the integrated QMS signal for a particular m/z value, in this case m/z=29, $B(29)$. We empirically link the calibration constant $k_{\mathrm{CO}}$ and the CO column density ($N_{\mathrm{CO}}$) using the pure TPD experiments:
\begin{equation}\label{eqn:QMS3}
k_{CO} = \frac{B(29)}{N(\mathrm{CO})}
\end{equation}
and obtain a $k_{\mathrm{CO}}$ value of $8.87~\times~10^{-10}$. As our electron bombardment experiments are performed in a separate chamber, we repeat this process with an unirradiated H$_2$O:CO:NH$_3$ ice and obtain a $k_{\mathrm{CO}}$ value of $1.3~\times~10^{-10}$.

Both N$_{2}$ and CO are diatomic molecules of relatively similar sizes, and we therefore assume that the molecules exhibit similar desorption and QMS detection behavior in the set-up. We can write the equation to obtain  $N_\mathrm{N_2}$ in terms of ratios of known values for CO and N$_2$ and the calibration constant for CO, $k_{CO}$.  Following \citet{martin2015uv,simon2023entrapment}, B(m/z) is  described as:
\begin{equation}\label{eqn:QMS}
\begin{split}
B(m/z) = k_{QMS}~\cdot~\sigma^{+}(x)~\cdot~N(x)\\
\cdot~I_{F}(x)~\cdot~F_{F}(x)~\cdot~S(m/z)
\end{split}
\end{equation}
where $m$ is the mass of the species, $z$ is the charge of the species, $\sigma^{+}(x)$ is the ionization cross section (2.516 mol and 2.508 mol for CO and N$_{2}$, respectively), $I_{F}(x)$ is the fractionation of the molecule (0.949 and 0.933 for CO and N$_{2}$), $F_{F}(x)$ is the pattern of fractionation (1 for both for CO and N$_{2}$), and $S(m/z)$ is the QMS' sensitivity (3.17~$\times$~10$^{14}$ and 3.10~$\times$~10$^{14}$ for CO and N$_{2}$ from \citet{simon2023entrapment}).

To finally extract the N$_{2}$ column densities presented in this work, we first fit a linear baseline and then numerically integrate the peaks under the m/z=30 curve in the range from the irradiation temperature up to 185~K, i.e. beyond the complete desorption of water ice, to obtain $B{(30)}$. We note that in some of the experiments we see an additional desorption feature beyond 200~K. These peaks likely correspond to material desorbing from elsewhere within the chamber, though another possible origin is ammonium salts, which can form from NH$_3$ interactions with water or organic contaminants \citep{schutte2003origin}.  We combine and rearrange Equations~\ref{eqn:QMS3} and \ref{eqn:QMS} to determine our final column density of N$_{2}$ as:
\begin{equation}\label{eqn:QMS2}
\begin{split}
N(\mathrm{N_{2}}) =\frac{B(30)}{k\mathrm{_{CO}}}~\cdot~\frac{\sigma^{+}(\mathrm{CO})}{\sigma^{+}(\mathrm{N_{2}})}~\cdot~\frac{I_{F}(\mathrm{CO})}{I_{F}(\mathrm{N_{2}})}\\~\cdot~\frac{F_{F}(\mathrm{CO})}{F_{F}(\mathrm{N_{2}})}~\cdot~\frac{S(29)}{S(30)}
\end{split}
\end{equation}

The uncertainty of the N$_{2}$ column density is estimated based on the CO band strength uncertainty of 20\% and the QMS CO to N$_{2}$ calibration uncertainty ($\sigma_{QMS}$) of 12\% (as previously reported by \citet{simon2023entrapment}):
\begin{equation}
   \sigma_{N\mathrm{_{N_{2}}}} = \sqrt{(\sigma_{N\mathrm{_{IR(^{13}CO)}}})^2+(\sigma\mathrm{_{QMS}})^2},
\end{equation}
which results in a N$_2$ column density uncertainty of $\sim$24\%.

We also present the CO TPD experiments after CO$_2$ irradiation. However, the m/z=29 signal is contaminated from CO$_{2}$ desorbing and fractionating into CO. In the displayed curves, we attempted to remove the contamination, but this is not overly accurate. 

\begin{figure*}
\centering
  \includegraphics[width=18cm]{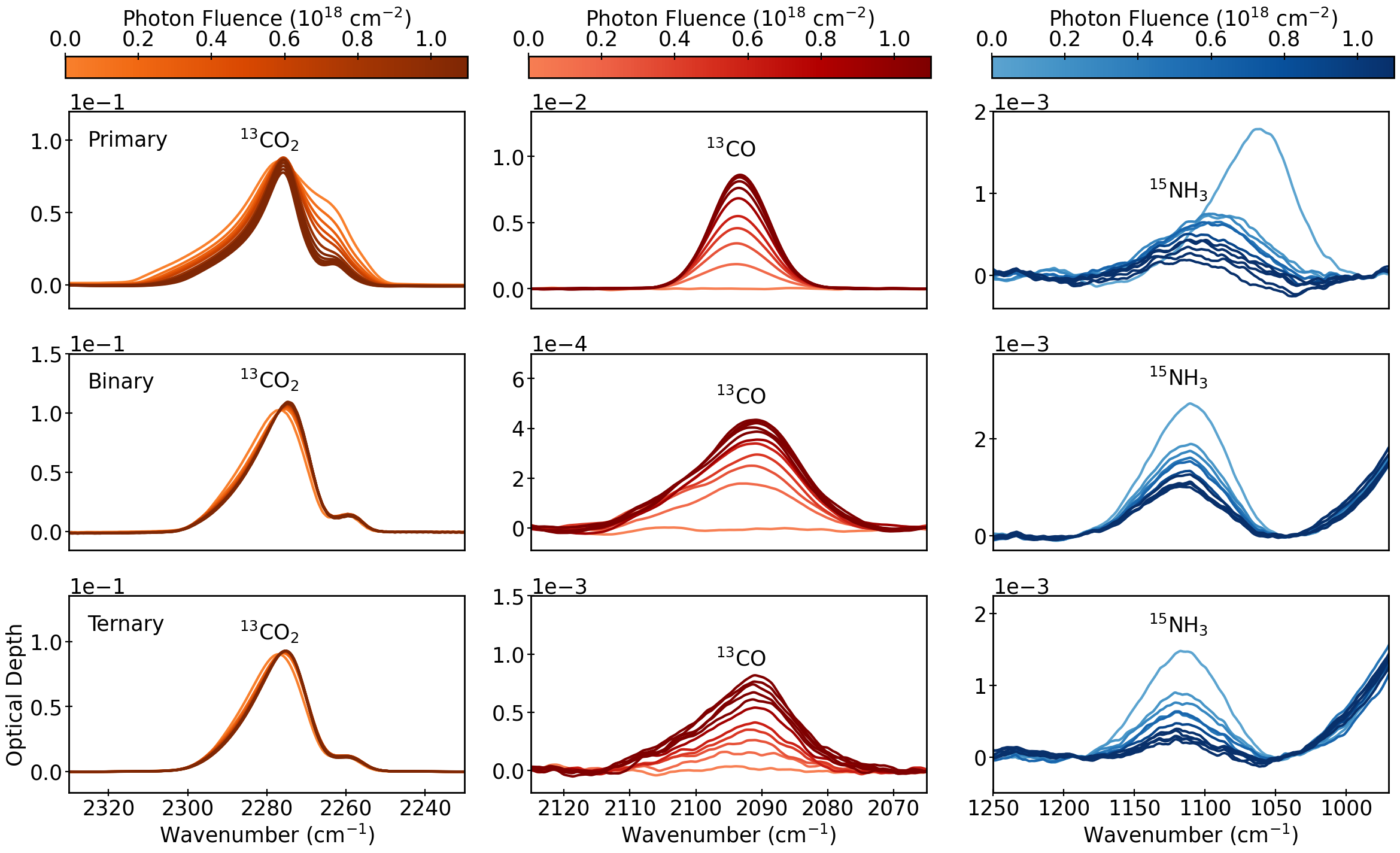}
  \caption{The infrared spectra of various ices throughout UV photolysis at 10~K. Left:CO$_{2}$; Middle:CO. Right:NH$_{3}$. Top Row: Pure Ices; Middle Row: Water-rich Binary Ices; Bottom Row: Water-rich ternary ices.}
  \label{fgr:Change_In_Peak}
\end{figure*}

\section{Results and Discussion} \label{sec:results}

\subsection{CO and N$_{2}$ Production in Undiluted CO$_{2}$ and NH$_{3}$ Ices}\label{subsec:Primary}

\begin{figure}
\centering
  \includegraphics[width=8cm]{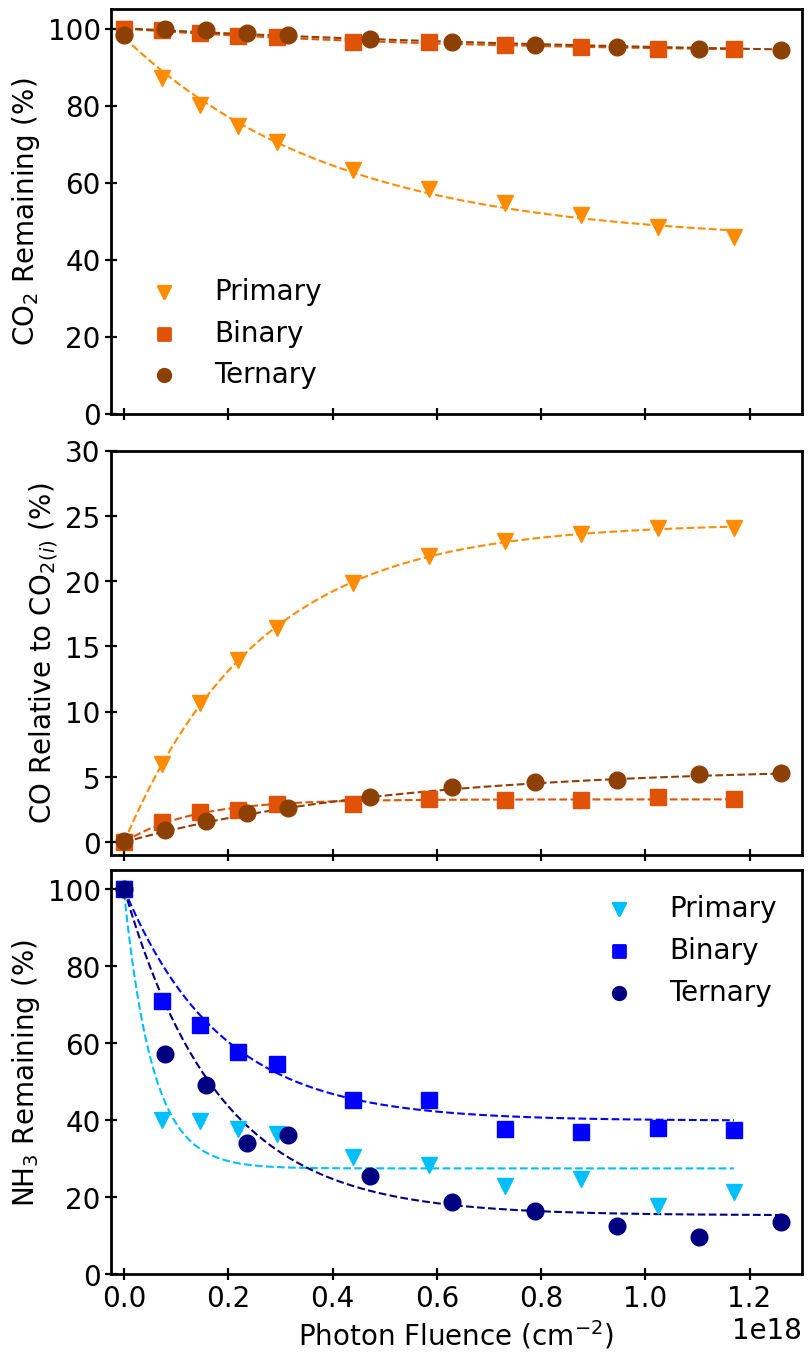}
  \caption{The temporal behavior of CO$_{2}$ (top panel) CO (middle panel) and NH$_{3}$ (bottom panel) following UV irradiation of various ice compositions at 10~K. The error bars in this plot only reflect the spectral uncertainty and are smaller than the markers.}
  \label{fgr:Temporal_Behaviour_PBT_SC}
\end{figure}

We begin by measuring the CO$_{2}$-CO and NH$_{3}$-N$_{2}$ conversion in pure ices at 10 K. Beginning with the CO$_{2}$-CO chemistry, Figure~\ref{fgr:Change_In_Peak} (top row) shows the  undiluted ice IR spectra throughout UV irradiation. 
Upon exposure to UV radiation, the CO$_{2}$ peaks decay while a CO peak appears around 2090~cm$^{-1}$  \citep{gerakines1994infrared}. Figure~\ref{fgr:Temporal_Behaviour_PBT_SC} (upper and middle panels) presents the resulting CO$_2$/CO$_{2(i)}$ and CO/CO$_{2(i)}$ column density ratios.
The CO column density increases rapidly at early irradiation times, and begins to plateau around a fluence of 5~$\times~10^{17}$ photons~cm$^{-2}$. Similarly, the CO$_{2}$ column density initially decreases at a rapid rate and then begins to plateau around the same UV fluence. 

At the end of the experiment, approximately half of the consumed CO$_{2}$ ice has formed CO, and since about half of the initial CO$_{2}$ was consumed, the final CO: CO$_{2}$ mixing ratio is $\sim$50\%. We cannot account for all of the consumed CO$_2$ not converted into CO (only a fraction shows up as more complex oxides), and we suspect that photodesorption may play a role for this thin ice \citep{oberg2007PHOTODESORPTION} We note that our CO yield is a lower limit as we detected an elevated m/z=29 signal during irradiation, indicating that some CO may be photodesorbing from the ice \citep{oberg2007PHOTODESORPTION,oberg2009photodesorption}. This is similar to what was previously reported by \citet{martin2015uv}. The measured initial and final CO$_2$ column densities, the final CO column density, and the relevant ratios are shown in Table \ref{tbl:Summary_of_Results}. The uncertainty in the reported ratios also include results from repeat experiments, which resulted in variations in CO$_2$-CO conversion efficiencies of 20 \%, which we add in quadrature to error of individual column density calculations (see Appendix~\ref{app:Repeat}). 
We also repeated this experiment using $^{12}$CO$_{2}$ and found no significant difference in $^{12}$CO yield compared to the equivalent 13-C labeled experiment (Appendix~\ref{app:12CO}). \citet{martin2015uv} also did not observe any significant difference between $^{12}$CO$_{2}$ and $^{13}$CO$_{2}$ photolysis. 

\begin{figure*}[ht!]
\centering
  \includegraphics[width=18cm]{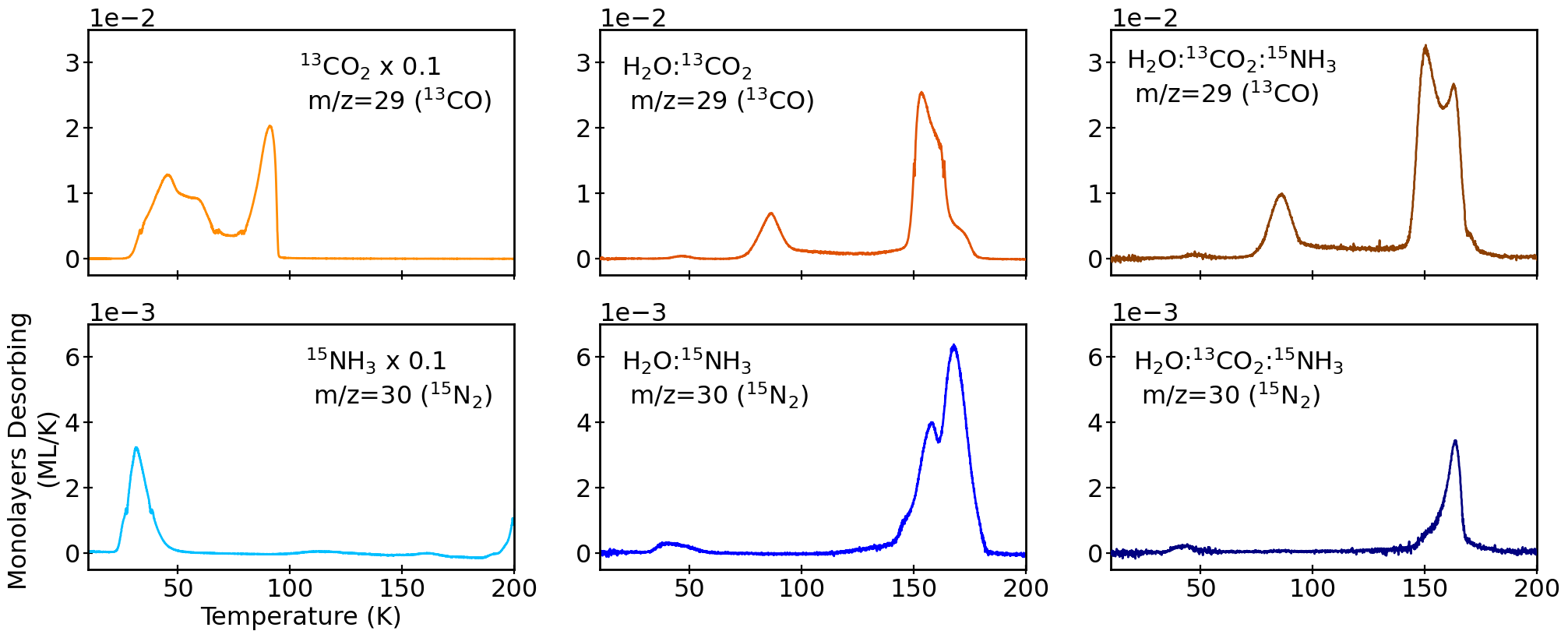}
  \caption{The TPD of $^{13}$CO (m/z=29 top row) and $^{15}$N$_{2}$ (m/z=30 bottom row) following UV irradiation of various ices at 10~K. Left column: primary ices; Middle column: water-rich binary ices; Right column: water-rich ternary ices. Note that the high-temperature peak around 200~K in some TPD traces (bottom left panel) are presumed to correspond to either codesorption of hypervolatiles with ammonium salts or to desorption from another part of the chamber. These peaks are not included in the abundance calculation to avoid over-reporting. Some of the TPD curves have been scaled for readability, which is listed on the individual panel.}
  \label{fgr:TPD_SC_PBT}
\end{figure*}

Figure~\ref{fgr:Change_In_Peak} also shows the decay in the NH$_{3}$ upon UV exposure of the undiluted NH$_{3}$ ice. Figure~\ref{fgr:Temporal_Behaviour_PBT_SC} shows a substantially faster destruction of NH$_{3}$ compared to that of CO$_{2}$. This difference is apparent in the fitted destruction cross sections of (18~$\pm$~6)~10$^{-18}$~cm$^2$ and (2.5~$\pm$~0.6)~10$^{-18}$~cm$^2$, for NH$_3$ and CO$_2$ respectively. This difference is consistent with the order of magnitude lower VUV absorption cross section of CO$_{2}$ ice compared to NH$_{3}$ ice \citep{cruz2014vacuum2,cruz2014vacuum3,cruz2014vacuum1}. 

We use the TPD experiments presented in Figure~\ref{fgr:TPD_SC_PBT} to quantify $^{15}$N$_{2}$ production (m/z=30, lower panels), but also show the $^{13}$CO traces (m/z=29, upper panels) for comparison. Because the TPD curves have been calibrated to show ML/K, the peak area corresponds to the number of monolayers desorbed, and hence produced during the experiment. We note that we expect both N$_{2}$ and CO to desorb around 30~K, which is consistent with the first peaks in the single-component ice experiments. The later peaks correspond to desorption during CO$_{2}$ and NH$_{3}$ crystallization and desorption, respectively \citep{collings2004laboratory}, due to entrapment in these ices \citep{simon2023entrapment}. We report the integrated N$_{2}$ TPD curve between 20~K and 185~K in Table~\ref{tbl:Summary_of_Results}. Similarly to for CO$_2$ and CO, the uncertainties in the N$_2$/NH$_3$ ratios include an experimental variation in the conversion efficiency from repeat experiments of 20~\%, as shown in Appendix~\ref{app:Repeat}.

\begin{deluxetable*}{c||ccc|ccc|ccc}[ht]
\tabletypesize{\scriptsize} 
\tablewidth{0pt}
\centering
%\tablenum{1}
\tablecaption{Summary of the CO and N$_{2}$ formation from  irradiation of various ices CO$_2$ and NH$_3$-containing ices.}
\label{tbl:Summary_of_Results}
\tablehead{
\colhead{Parameter} &\colhead{Primary} &\colhead{Binary} &\colhead{Ternary}&\colhead{Ternary} &\colhead{Ternary} &\colhead{Ternary}&\colhead{Primary$^c$} &\colhead{Binary$^c$} &\colhead{Ternary$^c$}\\
\colhead{} &\colhead{10~K} &\colhead{10~K} &\colhead{10~K}&\colhead{30~K} &\colhead{50~K} &\colhead{100~K}&\colhead{10~K} &\colhead{10~K} &\colhead{10~K}\\
\colhead{} &\colhead{UV} &\colhead{UV} &\colhead{UV}&\colhead{UV} &\colhead{UV} &\colhead{UV}&\colhead{e$^-$} &\colhead{e$^-$} &\colhead{e$^-$}}
\startdata
CO$_{2(i)}$	(ML)$^a$	&	22	$\pm$	5	&	17	$\pm$	4	&	15	$\pm$	3	&	16	$\pm$	3	&	12	$\pm$	2	&	13	$\pm$	3	&	$\sim$~17	&	$\sim$~9	&	$\sim$~12	\\
$\Delta$CO$_{2}$	(ML)$^a$	&	12	$\pm$	5	&	0.9	$\pm$	0.4	&	0.9	$\pm$	0.4	&	1.2	$\pm$	0.5	&	1.0	$\pm$	0.4	&	3	$\pm$	1	&	$\sim$~5	&	$\sim$~2.2	&	$\sim$~1.7	\\
$\Delta$CO	(ML)$^a$	&	6	$\pm$	1.1	&	0.40	$\pm$	0.08	&	0.8	$\pm$	0.2	&	0.6	$\pm$	0.1	&	0.7	$\pm$	0.1	&	1.0	$\pm$	0.2	&	$\sim$~8	&	$\sim$~0.3	&	$\sim$~0.6	\\
$\Delta$CO/CO$_{2(i)}$	(\%)$^b$	&	25	$\pm$	7.1	&	2.3	$\pm$	0.7	&	5	$\pm$	2	&	4	$\pm$	1	&	6	$\pm$	2	&	8	$\pm$	2	&	$\sim$~44	&	$\sim$~3.2	&	$\sim$~5	\\
$\Delta$CO/$\Delta$CO$_{2}$	(\%)$^b$	&	48	$\pm$	22	&	44	$\pm$	20	&	92	$\pm$	42	&	47	$\pm$	21	&	67	$\pm$	31	&	30	$\pm$	14	&	$\sim$~100	&	$\sim$~14	&	$\sim$~35	\\
$\Delta$CO/CO$_{2(f)}$	(\%)$^b$	&	51	$\pm$	14.7	&	2.5	$\pm$	0.7	&	6	$\pm$	2	&	4	$\pm$	1	&	7	$\pm$	2	&	10	$\pm$	3	&	$\sim$~62	&	$\sim$~4	&	$\sim$~6	\\
$\Delta$CO/H$_{2}$O	(\%)$^b$	&		N/A		&	0.4	$\pm$	0.1	&	0.9	$\pm$	0.2	&	0.5	$\pm$	0.1	&	0.6	$\pm$	0.2	&	0.8	$\pm$	0.2	&		N/A		&	$\sim$~0.5	&	$\sim$~0.5	\\
\hline
NH$_{3(i)}$	(ML)$^a$	&	7	$\pm$	1.5	&	9	$\pm$	2	&	5	$\pm$	1	&	5	$\pm$	1	&	4.5	$\pm$	0.9	&	5	$\pm$	1	&	$\sim$~48	&	$\sim$~9	&	$\sim$~5	\\
$\Delta$NH$_{3}$	(ML)$^a$	&	5	$\pm$	2.1	&	6	$\pm$	2	&	4	$\pm$	2	&	4	$\pm$	2	&	4	$\pm$	1	&	4	$\pm$	2	&	$\sim$~16	&	$\sim$~5	&	$\sim$~4	\\
$\Delta$N$_{2}$	(ML)$^a$	&	0.3	$\pm$	0.1	&	0.13	$\pm$	0.05	&	0.05	$\pm$	0.02	&	0.04	$\pm$	0.02	&	0.05	$\pm$	0.02	&	0.11	$\pm$	0.04	&	$\sim$~4	&	$\sim$~0.7 &	$\sim$~0.20	\\
$\Delta$N$_{2}$/NH$_{3(i)}$	(\%)$^b$	&	4	$\pm$	2	&	1.0	$\pm$	0.5	&	1.0	$\pm$	0.4	&	0.7	$\pm$	0.3	&	1.1	$\pm$	0.4	&	2.0	$\pm$	0.8	&	$\sim$~9	&	$\sim$~7	&	$\sim$~4	\\
$\Delta$N$_{2}$/$\Delta$NH$_{3}$	(\%)$^b$	&	6	$\pm$	3	&	2	$\pm$	1	&	1.2	$\pm$	0.6	&	0.8	$\pm$	0.4	&	1.3	$\pm$	0.7	&	2	$\pm$	1.2	&	$\sim$~20	&	$\sim$~12	&	$\sim$~6	\\
$\Delta$N$_{2}$/NH$_{3(f)}$	(\%)$^b$	&	21	$\pm$	8	&	4	$\pm$	1	&	7	$\pm$	3	&	6	$\pm$	2	&	6 $\pm$	2	&	17	$\pm$	7	&	$\sim$~15	&	$\sim$~19	&	$\sim$~11	\\
$\Delta$N$_{2}$/H$_{2}$O	(\%)$^b$	&		N/A		&	0.13	$\pm$	0.05	&	0.05	$\pm$	0.02	&	0.03	$\pm$	0.01	&	0.04	$\pm$	0.02	&	0.09	$\pm$	0.04	&		N/A		&	$\sim$~0.7	&	$\sim$~0.18	\\
\enddata
\tablecomments{All final column densities are based on the last IR spectrum acquired at the end of irradiation. $\Delta$ is used to denote the change in column density over the entirety of the experiment.\\ $^a$Reported uncertainties only reflect spectral uncertainty from each individual experiment and do not reflect experimental error.\\ $^b$Reported uncertainties reflect both spectral uncertainty and experimental error from repeat experiments.\\$^c$The column densities have a larger uncertainty due to longitudinal modes from reflection-absorption mode \citep{mate2008ices}. Our uncertainties are at least 50\% but they could be larger.}
\end{deluxetable*}

Table \ref{tbl:Summary_of_Results} shows that  about 12\% of the destroyed NH$_{3}$ ice formed N$_{2}$ (taking into account that two NH$_3$ are required to form one N$_2$). This corresponds to 8\% of the initial NH$_{3}$ ice. The final N$_2$:NH$_3$ mixing ratio is 21\%. These values are all lower compared to the equivalent ones characterizing the CO conversion yield. This is not surprising considering the number of dissociation and reaction steps that are needed to convert NH$_3$ to N$_2$ compared to converting CO$_2$ to CO. Indeed the majority of consumed NH$_3$ is likely accounted for by photodissociation and the formation of other products in the NH$_{3}$ ice, such as ammonium salts \citep{martin2018uv}; there are some IR signatures in the irradiated NH$_3$ spectra that may be due to ammonium salts (see Appendix~\ref{app:IR_Spectra}) but analyzing this in detail is beyond the scope of this work.

\subsection{CO and N$_{2}$ Production in Water-rich CO$_{2}$ and NH$_{3}$ Binary Ices}\label{subsubsec:Binary}

To better constrain our reaction in a more astrophysically relevant ice, we next consider UV photodestruction in H$_{2}$O:CO$_{2}$ and H$_{2}$O:NH$_{3}$ binary ices. This also serves to constrain the matrix effect on the hypervolatile production from CO$_{2}$ and NH$_{3}$. The infrared spectra of these ices are in the middle row of Figure~\ref{fgr:Change_In_Peak}. Similarly to the undiluted ice experiments, the destruction curves of CO$_{2}$ and NH$_{3}$, and  the formation curve of CO are shown in Figure~\ref{fgr:Temporal_Behaviour_PBT_SC}, and the initial and final reactant column densities, the final hypervolatile column densities and the relevant ratios are presented in Table \ref{tbl:Summary_of_Results}. As for undiluted ices, the reported column densities are presented with uncertainties based on single experiments, while the ratios used to estimate formation efficiencies of hypervolatiles also include a 20\% experimental variation error term in the reported uncertainties. 

In the water-rich CO$_{2}$-containing binary ice, the top and middle panels of Figure~\ref{fgr:Temporal_Behaviour_PBT_SC} show a dramatic decrease in both CO$_{2}$ destruction and CO formation compared to the undiluted ice.  Table~\ref{tbl:Summary_of_Results} reports the yield, and  only 3~$\%$ of the initial CO$_{2}$ formed CO which accounts for about one third of the CO$_{2}$ consumed. This is a decrease in $\Delta$CO/CO$_{2(i)}$ of almost 90\% compared to the undiluted CO$_{2}$ ice, but the percentage of the consumed CO$_{2}$ that is converted into CO remains consistent. The final mixing ratios for CO:CO$_2$ and CO:H$_2$O are $\sim$3 and 0.4\%, respectively.

Figures~\ref{fgr:Change_In_Peak} and \ref{fgr:Temporal_Behaviour_PBT_SC} show that in the water-rich NH$_{3}$-containing binary ice, the difference in NH$_{3}$ destruction between the undiluted ice and the water-rich binary ice is more modest compared to that of CO$_{2}$.  This difference suggests that NH$_{3}$ photodissociation may be less impacted by the cage effect than CO$_2$ photodissociation. In previous studies amorphous water ices have displayed strong cage effects that prevent dissociation fragments  from escaping their position within the film \citep{braden2001solvent,mcneill2012organics}, enhancing recombination and therefore reducing the net photodissociation yield. This has been seen for several volatile species  \citep{bar1985trapping,bar1987amorphous,bar1988trapping}. We suggest that the relative caging of CO$_2$ dissociation fragments (CO and O) is likely more efficient than of NH$_3$ dissociation fragments (especially H), since the former are considerably larger. The corresponding TPD experiments in Figure~\ref{fgr:TPD_SC_PBT} show a consistent pattern with smaller peaks for both CO (m/z=29) and N$_{2}$ (m/z=30) compared to the undiluted ices (note that the latter are plotted at a $\times$0.1 scale). Note that when the hypervolatiles are in water-rich ices, they are mostly entrapped in the water matrix and desorb around water's desorption temperature of $\sim$150~K \citep{collings2004laboratory}.  The substantial reduction of N$_2$ formation even as NH$_3$ dissociation only slightly decreased indicates that a larger proportion of NH$_3$ dissociation events leads to other products in a water ice matrix compared to the undiluted ice.

Quantitatively, the  N$_{2}$ yield in the water-rich binary ices is one third to one quarter that of undiluted ices, dependent on whether N$_2$ production is compared to initial, consumed, or final NH$_3$ (Table~\ref{tbl:Summary_of_Results}). In the binary ice, $\sim$2~$\%$ of the initial NH$_{3}$ ice formed N$_{2}$, corresponding to 4~$\%$ of the destroyed NH$_{3}$ ice (note that the numbers in the Table are half of this since these are abundance ratios). We note that while for CO$_2$ there is no change in the ratio between produced CO and consumed CO$_2$ (i.e. the product branching ratio) when comparing the undiluted and binary ice experiments, we do see a clear decrease in this ratio for N$_2$ and consumed NH$_3$. This can be understood when considering that while CO is a direct photodissociation product of CO$_2$, N$_2$ formation requires the diffusion of two NH$_3$ dissociation fragments, and this may be hindered by cage effects in the water ice matrix. i.e. while the water ice may not be very good at inhibiting H diffusion, it may efficiently reduce the diffusion and recombination of larger N-containing fragments. The final N$_2$:NH$_3$ and N$_2$:H$_2$O mixing ratios are $\sim$4 and 0.1\%, respectively.

\subsection{CO and N$_{2}$ Production in H$_2$O:CO$_{2}$:NH$_{3}$ Ternary Ices}\label{subsubsec:Ternary}

We next studied H$_{2}$O:CO$_{2}$:NH$_{3}$ ternary ices where the ice mixing ratio of 100:20:5 mimics the typical composition of the water-rich protostellar ice phase \citep{boogert2015observations}. Figure~\ref{fgr:Change_In_Peak} shows that spectroscopically the 10~K ternary photochemistry experiment appears similar to the binary ones, and the CO$_2$ and NH$_3$ dissociation curves, and CO formation curves also resemble those of binary ices. The final CO/CO$_{2(i)}$ column density ratio is, however, about a factor of two higher in the ternary ice compared to the binary ice (Figure~\ref{fgr:Temporal_Behaviour_PBT_SC}). This may be explained by the presence of NH$_{3}$, which could be softening the water-ice matrix, and hence allowing for greater CO$_{2}$ destruction by reducing the cage effect for CO$_{2}$ in the water ice. We tested this hypothesis by running an additional ternary ice experiment with twice as much NH$_3$, and found that this increased both CO$_2$ destruction and CO formation further, tentatively confirming the proposed explanation (Appendix~\ref{app:TwiceAmmonia}). By contrast the N$_{2}$ production in the ternary ice is slightly depressed compared to the binary ice, though not significantly so. The presence of CO$_{2}$ in the ice may open up additional chemical pathways for NH$_{3}$ photodissociation fragments, including for efficient formation of ammonium salts, which would also support the increase in m/z=30 signal around 200~K in both Figure~\ref{fgr:TPD_SC_PBT} and Appendix~\ref{app:TPDs}.

\begin{figure}
\centering
  \includegraphics[width=8cm]{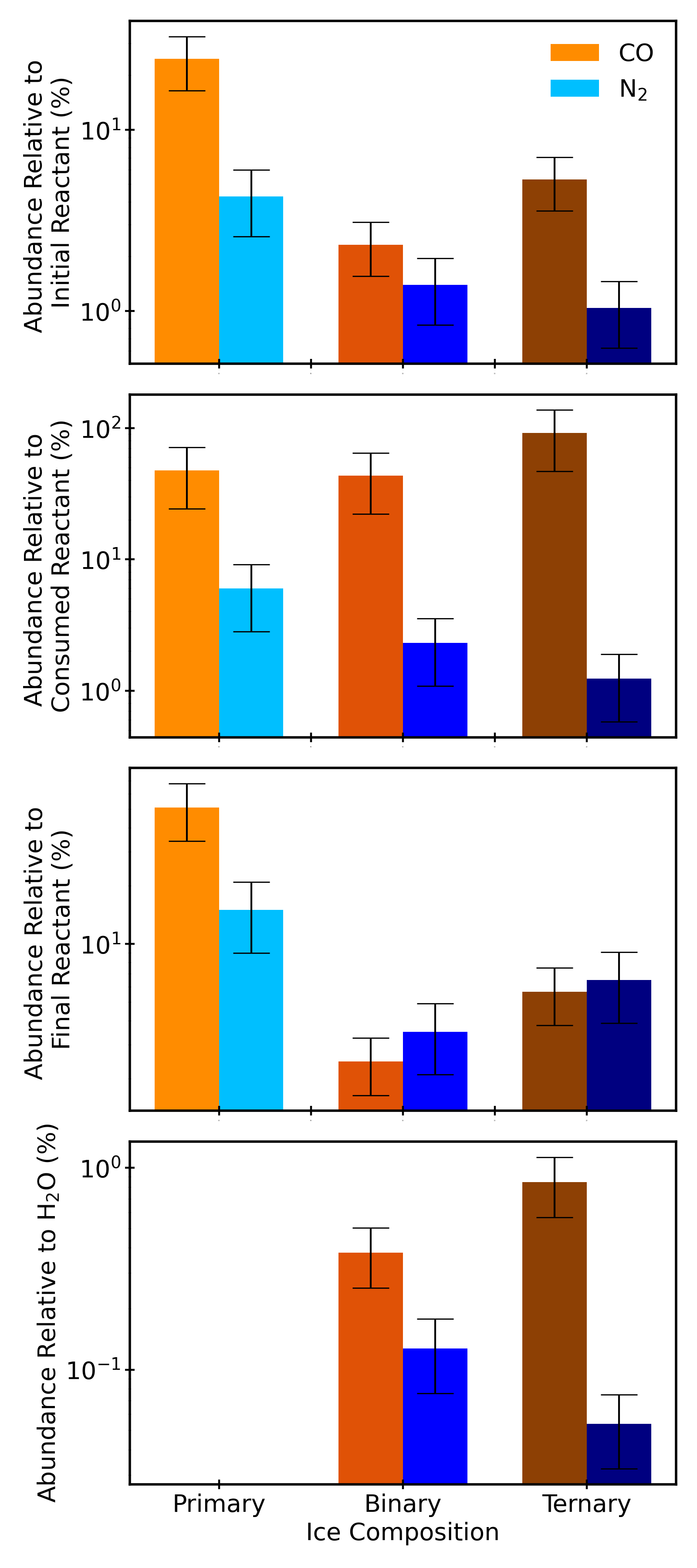}
  \caption{A summary of the CO and N$_{2}$ formed following irradiation of various ices at 10~K. The initial reactant for CO is the number of CO$_{2}$ monolayers before irradiation, and the initial reactant for N$_{2}$ is the number of NH$_{3}$ monolayers before irradiation. The panels show the hypervolatile yield relative to the initial CO$_{2}$ and NH$_{3}$ abundance (the top panel), to the  consumed CO$_{2}$ and NH$_{3}$ (the second panel), to the final CO$_{2}$ and NH$_{3}$ abundance (the third panel) and to the abundance of water in the ice (the bottom panel).}
  \label{fgr:Histogram_Composition}
\end{figure}

Figure~\ref{fgr:Histogram_Composition} summarizes the different CO and N$_2$ production efficiencies across the undiluted, binary and ternary experiments at 10 K. The top panel shows N$_2$ and CO yields compared to initial reactants, and the two notable features are 1: that the CO yield is a factor of 2--10 higher than N$_2$ (with the smallest differences in binary ices), and 2: that the yields decrease by up to an order of magnitude between pure and water-rich ices. The second panel from the top illustrates the product branching ratios by plotting CO and N$_2$ yields compared to consumed CO$_2$ and NH$_3$, and we see that these are stable for CO, and decrease with ice complexity for N$_2$. The third panel from the top
displays the abundance of hypervolatiles formed with respect to the final reactant abundance in order to show the CO:CO$_{2}$ and N$_{2}$:NH$_{3}$ mixing ratios that we would expect in processed interstellar and cometary ices. These show similar trends to the top panel, but we note that in the case of N$_{2}$, our yield with respect to the final NH$_{3}$ column density is slightly larger in the complex ternary ice than in the water-rich binary ice. 
Finally, the bottom panel displays the CO and N$_{2}$ yields with respect to the amount of water present in the ices to provide a more direct comparison to the reported CO and N$_{2}$ abundances previously reported in comets. 

In addition to our fiducial ternary experiment, we also ran one experiment with a reduced amount of NH$_3$, similar to what is seen in most comets, to account for the possibility of lower initial NH$_3$ abundance in the Solar Nebula compared to the typical protostellar environment. In this experiment, when we start with a ternary ice containing 1\% NH$_3$ with respect to water, we form approximately 0.009 ± 0.004 \% N$_2$ with respect to water, as shown in Appendix~\ref{app:TwiceAmmonia}. The factor of five reduction suggests a linear relationship between initial NH$_3$ abundance and produced N$_2$.

\subsection{Irradiation Temperature Comparison} \label{subsec:Temperature}

\begin{figure}
\centering
          \includegraphics[width=8cm]{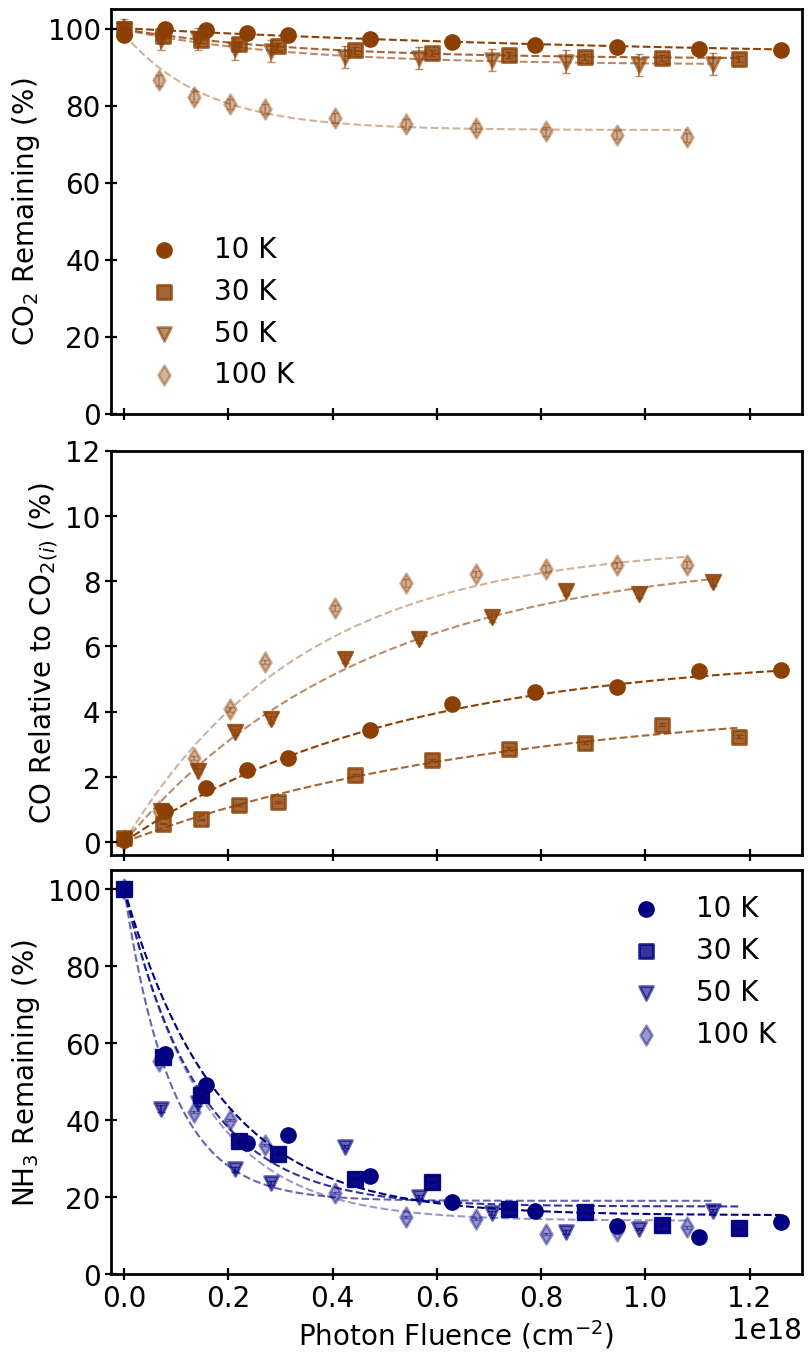}
  \caption{The temporal behaviour of CO$_{2}$ (top panel) CO (middle panel) and NH$_{3}$ (bottom panel) following UV irradiation of ternary ices at various temperatures. The error bars in this plot only reflect the spectral uncertainty and are smaller than the markers.}
  \label{fgr:Temporal_Behaviour_SC_Temp}
\end{figure}

We next investigate the impact of temperature at which the H$_{2}$O:CO$_{2}$:NH$_{3}$ ice is irradiated on the hypervolatile production. The ternary mixture was irradiated at 10~K (Section~\ref{subsubsec:Ternary}), 30~K, 50~K, and 100~K for approximately the same amount of UV exposure of $\sim$1$\times$10$^{18}$ photons cm$^{-2}$ (see spectral time series in Appendix~\ref{app:IR_Spectra}). As 100~K is above the CO$_{2}$ desorption temperature, the initial CO$_{2}$ concentration reflects the entrapped CO$_2$. The top panels of Figure~\ref{fgr:Temporal_Behaviour_SC_Temp} show an increasing CO$_{2}$ destruction rate and increasing CO production rate with temperature. At 100~K, the increase in destruction is, however, higher than the increase in production, which can be explained by outgassing of CO and CO$_{2}$ at this elevated ice temperature. Table~\ref{tbl:Summary_of_Results} also shows an increasing CO/CO$_{2(i)}$ and CO/CO$_{2(f)}$ with temperature above 30 K, as well as a fairly stable branching ratio of $\sim$50\%, i.e. $\sim$50\% of the destroyed CO$_2$ is converted into CO.
Similar to CO$_2$, NH$_{3}$ destruction  rates increase somewhat with temperature. The N$_2$ formation yields also increase by a factor of $\sim$2 across the temperature experiments for both the N$_2$/NH$_{3, (i)}$ and N$_2$/NH$_{3, (f)}$ ratios.

\begin{figure}
\centering
  \includegraphics[width=8cm]{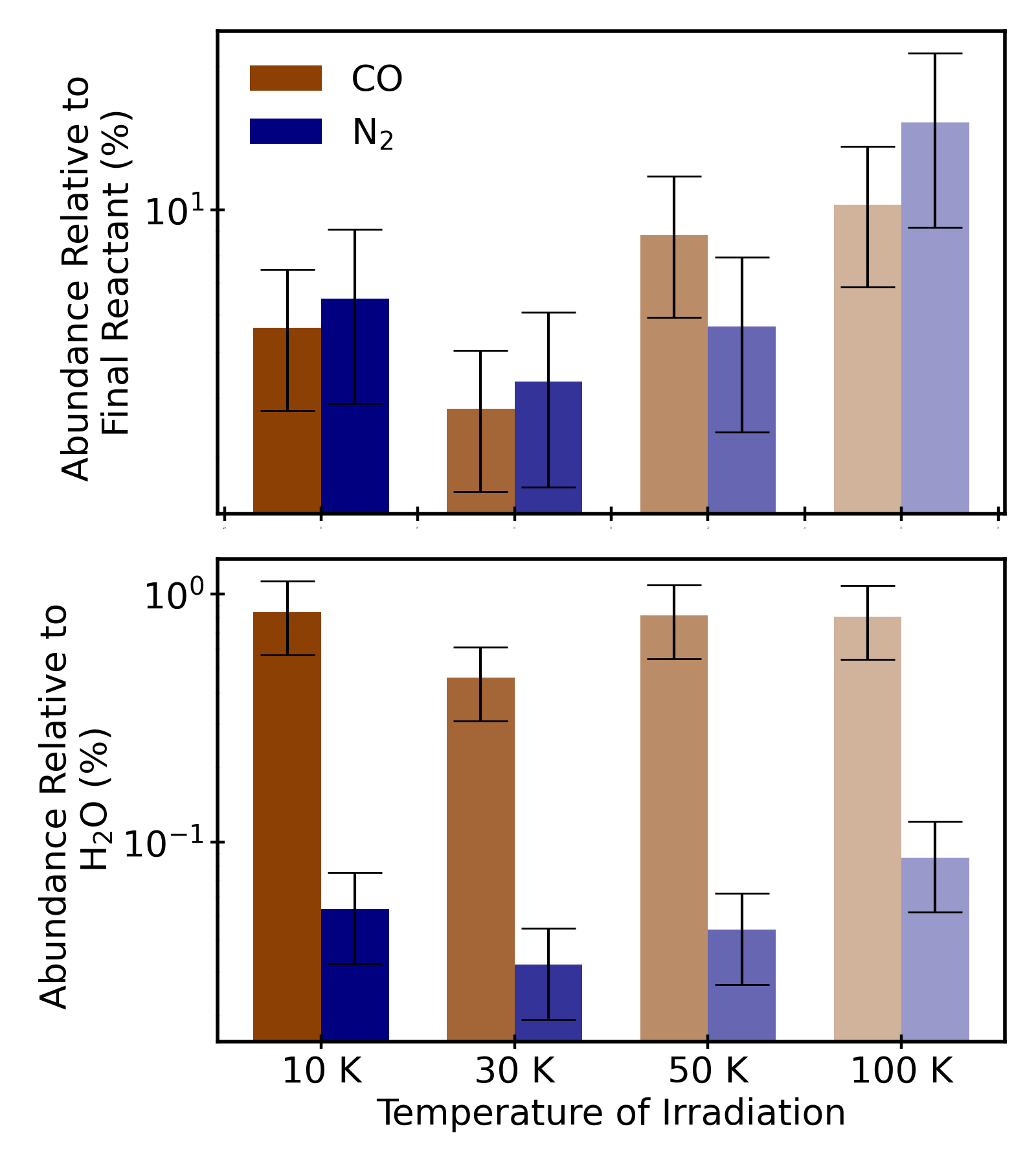}
  \caption{A summary of the CO and N$_{2}$ formed following irradiation of ternary ices at various temperatures. The reactant for CO is CO$_{2}$, and the reactant for N$_{2}$ is NH$_{3}$.}
  \label{fgr:Histogram_Temperature}
\end{figure}

Figure~\ref{fgr:Histogram_Temperature} shows the results most relevant experimental results for comparing with observations. The top panel shows the relative amount of CO and N$_{2}$ produced compared to the final amount of CO$_{2}$ and NH$_{3}$, respectively, i.e. the expected mixing ratios of reactants to products at steady-state. The second panel presents the final CO/N$_2$:water mixing ratios in the ice analogs. The final CO/CO$_{2}$ ratio increases with temperature beyond 30 K, which we attribute to more efficient CO$_{2}$ destruction at higher temperatures, due to less efficient cage effects as diffusion barriers become easier to overcome \citep{oberg2009formation}. The same trend is observed for N$_2$/NH$_3$, which can similarly be explained by reduced diffusion barriers for NH$_3$ dissociation fragments. There is a slightly lower CO yield at 30~K than at 10~K, which may be caused by desorption of the newly formed hypervolatile molecules at their sublimation temperature \citep{collings2003laboratory}. While the difference is not statistically significant, the same trend is also observed for N$_{2}$, which has a similar sublimation temperature. The final CO/H$_2$O and N$_2$/H$_2$O ratios also show a minimum at 30~K and then increase, but more modestly. Across the temperature regime the CO/H$_2$O and N$_2$/H$_2$O ratios are 50\% -- 90\% and 3\% -- 9\%, respectively, i.e. within the range observed for 10~K experiments. 

\subsection{Photons vs Electrons} \label{subsec:TertiaryTiger}

\begin{figure}
\centering
  \includegraphics[width=8cm]{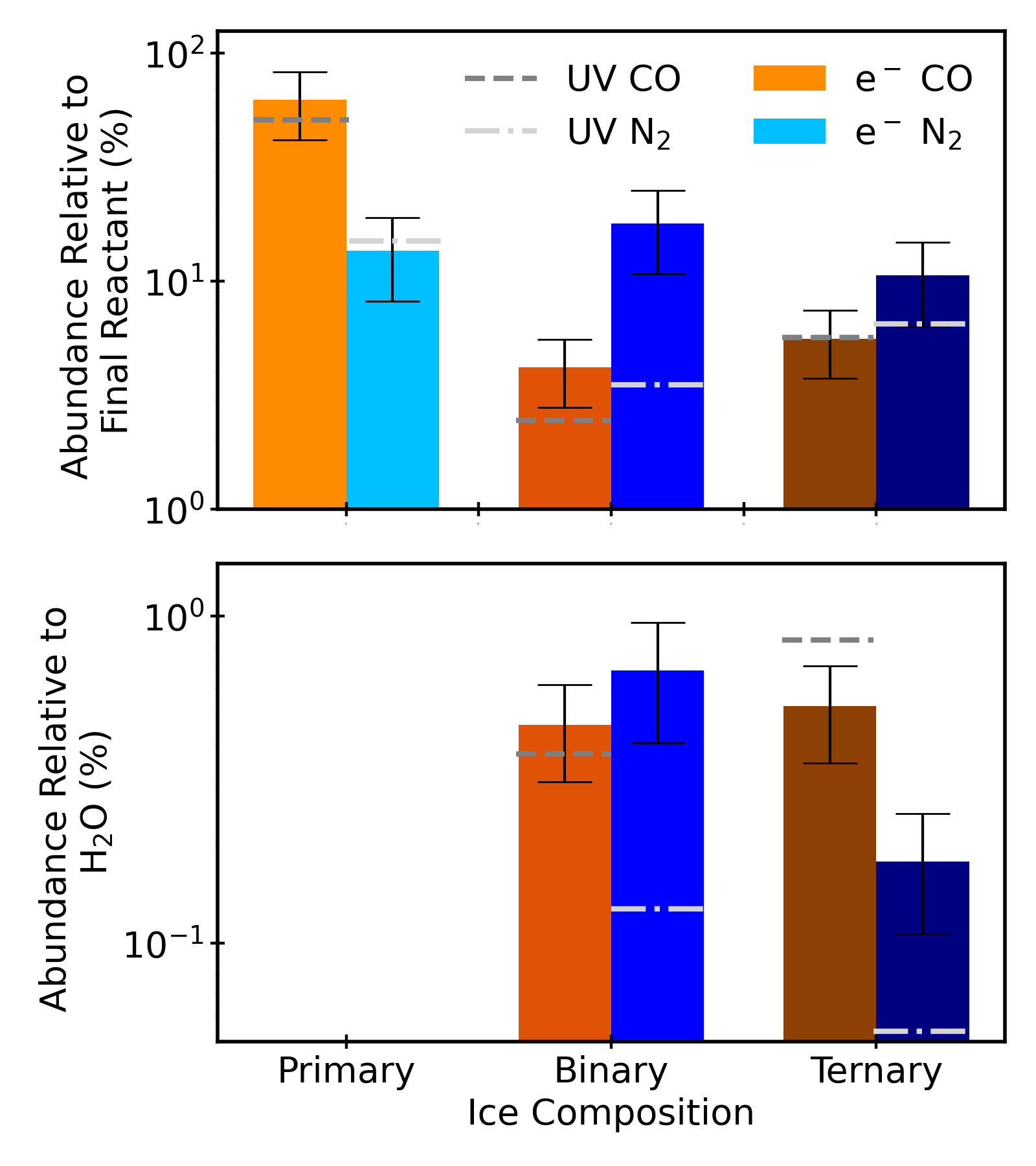}
  \caption{A summary of the CO and N$_{2}$ formed following electron bombardment of various ice compositions. The reactant for CO is CO$_{2}$, and the reactant for N$_{2}$ is NH$_{3}$. The horizontal dashed lines represent the CO and N$_{2}$ abundances reported following UV photolysis in Section~\ref{subsec:Primary}.}
  \label{fgr:Histogram_Electrons}
\end{figure}

Lastly, we investigated how the dissociation source can impact the production of hypervolatiles across the different ice compositions at 10~K.  To facilitate a comparison the ices were irradiated until approximately reaching a steady state in both kinds of experiments. The electron experiments did, however, receive an energy dose that was higher by a factor of a few (as discussed in Section~\ref{sec:Gas}), which may bias these experiments towards higher yields.  The relevant infrared and TPD data are shown in Appendices~\ref{app:IR_Spectra} and \ref{app:TPDs}, respectively, and the destruction and production curves in Appendix~\ref{app:SpaceTigerTemporalBehaviour}. Here we discuss the extracted final ice ratios and yields as presented in Table~\ref{tbl:Summary_of_Results} and Figure~\ref{fgr:Histogram_Electrons}.

Similarly to the UV photolysis experiments, the CO and N$_2$ yields (compared to initial reactant) are higher in the undiluted electron-bombarded ices compared to the water-rich mixtures (Table~\ref{tbl:Summary_of_Results}).  
Comparing the equivalent FUV and electron exposed experiments, CO and N$_{2}$ yields vary by factors up to 3 and 5, respectively, with the radiolysis experiments generally achieving higher yields in water-rich experiments (Fig. \ref{fgr:Histogram_Electrons}).
We note that it is somewhat surprising to see higher N$_2$ yields in the radiolysis experiments, since 2~keV electrons have enough energy to destroy N$_{2}$ while our FUV source does not \citep{cosby1993electron,song2023cross}. Also in the case of CO, we would expect more destruction in the electron bombardment experiment, since our FUV source only has weak emission below the Lyman $\alpha$ transition, and limited overlap with CO UV absorption bands \citep{piacentino2025survival}, though some indirect CO photodissociation may still occur.

The relatively large increase in the N$_2$ yield may point to more efficient NH$_3$ dissociation into smaller fragments (e.g. NH instead of NH$_2$) during radiolysis, facilitating N$_2$ formation, but confirming this hypothesis would require calculations beyond the scope of the present study. Perhaps most importantly for this study, the yields are still of the same order of magnitude, whether produced through photolysis or radiolysis.
This is not overly surprising, as previous work has found similar results from the radiolysis vs the photolysis of e.g. methanol ices \citep{mason2014electron,boamah2014low,oberg2016photochemistry}.
Finally we note that we record our electron bombardment IR spectra in reflection mode, which increases the uncertainties of the resulting column densities, since reflection-absorption mode band strengths are less well known and may change during the experiment due to changes in ice structure. We therefore report radiolysis yields as estimates.

\section{Astrochemical Implications}\label{sec:astrophys}

The previous section has demonstrated that CO and N$_{2}$ are formed from CO$_{2}$ and NH$_{3}$ dissociation across a range of astrochemically plausible conditions. In our pure experiments we find mixing ratios of CO/CO$_{2(f)}$ of 51--62\% and N$_{2}$/NH$_{3(f)}$ of 15--21\%. Across the water-rich experiments we find a range of yields and final mixing ratios such that the CO/CO$_{2(f)}$ ratio is 3--10\%, N$_{2}$/NH$_{3(f)}$ is $\sim$4--19\% , CO/H$_{2}$O is $\sim$0.4--0.9\%, and N$_{2}$/H$_{2}$O is $\sim$0.03--0.7\%. In the following sections we compare these ratios to those observed in comets, and those expected from entrapment to better understand the origin of cometary CO and N$_{2}$. 

\subsection{Comparison Between Photolysis/Radiolysis Experiments and Comet Observations}\label{subsec:expvscomet}

We compare our experimental ice mixing ratios with comet observations to explore whether photodissocation of mid-volatiles (CO$_2$ and NH$_3$) provide a plausible origin of cometary CO and N$_2$. A major assumption of this comparison is that our initial ice composition is similar to the primordial ice of the solar system, and hence that the lower NH$_3$/H$_2$O and CO$_2$/H$_2$O seen in comets \citep{mumma2011chemical,altwegg2019cometary} reflect consumption of parent molecules, through e.g. the kind of photolytic chemistry studied here, rather than a dramatically different initial set of conditions for the Solar Nebula compared to average protostellar systems. This assumption could certainly be contested and we revisit it below.

\begin{figure*}
\centering
  \includegraphics[width=18cm]{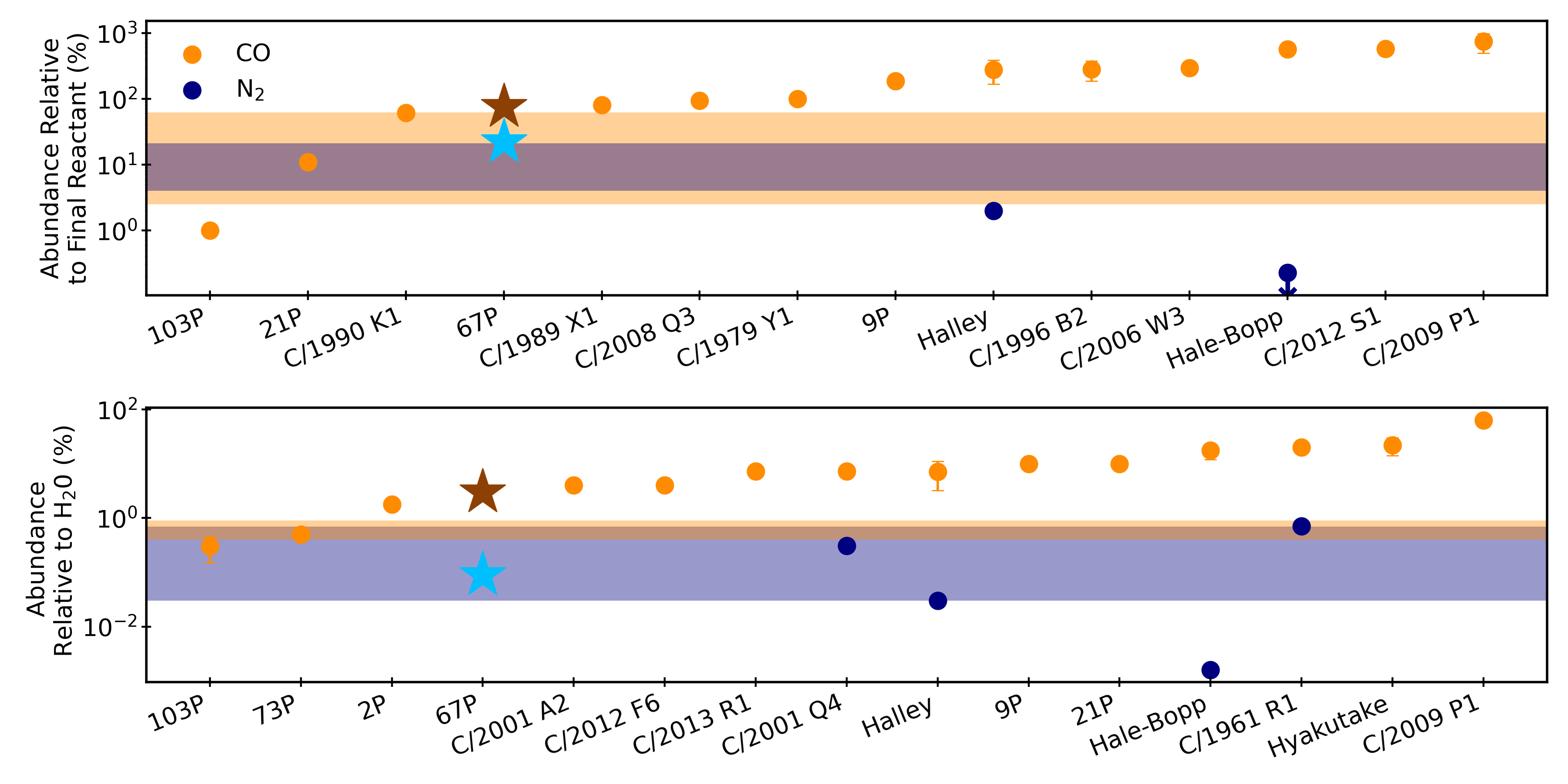}
  \caption{A summary of the CO and N$_{2}$ reported in various comets with respect to parent molecule (top) and with respect to water (bottom). The CO/CO$_{2}$ ratios are reported by \citet{pinto2022survey} and references within. N$_{2}$ abundances with respect to water are directly reported for three of these comets \citep{iro2003interpretation,le2015inventory,mckay2019peculiar}, and we calculated the N$_{2}$ abundances for C/2001~Q4 and C/1961~R1 from the N$_{2}$/CO and CO/H$_{2}$O ratios compiled by \citet{anderson2023n2}. For Comet 67P, we use the N$_2$ abundance reported by \citet{rubin2019elemental}, and the CO, CO$_2$, and NH$_3$ values reported by \citet{lauter2020gas}. For clarity, we exclude the results from Comet C/2016 R2, which report CO and N$_{2}$ abundances relative to water of (30800 $\pm$ 3500)~$\%$ and (1550 $\pm$ 370)~$\%$, respectively \citep{mckay2019peculiar}. Comet 67P markers are star shaped for emphasis. Shaded orange and blue regions represent the CO and N$_{2}$ abundances reported following photolysis and radiolysis, respectively. Downward-pointing arrows indicate upper limits on reported hypervolatile abundances.}
  \label{fgr:Compare_to_Comets}
\end{figure*}

Figure~\ref{fgr:Compare_to_Comets} (top panel) shows the CO/CO$_2$ and N$_2$/NH$_3$ abundances for comets compared to the range of CO/CO$_{2(f)}$ and N$_2$/NH$_{3(f)}$ abundances obtained in our experiments. The experimental N$_{2}$/NH$_{3(f)}$ ratios are comparable to those of 67P (marked with stars), but higher than the other two comets where this ratio is known \citep{le2015inventory,iro2003interpretation}, though considering uncertainties the difference is perhaps only significant for Hale-Bopp, which was unusually N$_2$ poor. Still the lower comet ratios may signal substantial outgassing of hypervolatiles, resulting in the smaller N$_{2}$ abundances reported in comets compared to our experimental results, or reflect lower initial NH$_3$ abundances. In the case of CO/CO$_{2(f)}$ ratios, we see consistency between our experimental results and three cometary abundances (103P, 21P, and C/1990~K1). However, we also see several cometary abundances exceeding those measured in the experiments, especially in our water-rich experiments. In particular, Comet 67P's reported CO/CO$_{2}$ ratio is similar to our yields from pure CO$_2$ photolysis and radiolysis experiments. Such a scenario may be closer to reality than may be initially assumed. Recent observations of CO ice in a protoplanetary disk shows it to be mainly mixed with CO$_2$ rather than H$_2$O \citep{Bergner2024}. However it is unclear if a similar scenario would have held for the Solar Nebula. Only $\sim$30\% of the CO outgassing in Comet 67P is correlated with CO$_2$ \citep{rubin2023volatiles}, suggesting that the majority of the CO is instead residing with H$_2$O \citep{rubin2023volatiles}. Hence for 67P a photolytic or radiolytic origin is only consistent for N$_2$. Similarly for the remaining comets the comparison is mixed with N$_2$/NH$_3$ ratios possibly explainable by photolysis/radiolysis, while the CO/CO$_2$ cometary ratios are too high. 

The bottom panel of Figure~\ref{fgr:Compare_to_Comets} shows the CO/H$_2$O and N$_2$/H$_2$O abundances for comets where at least the CO/H$_2$O ratio is known. Similar to the top panel, the most well-studied comet, Comet 67P, is marked by stars, and we overplot the range of CO and N$_2$ abundances obtained in our experiments. All cometary N$_{2}$ abundances with respect to water are consistent to or lower than the range of values produced from NH$_{3}$ photolysis or radiolysis in realistic protostellar ice mixtures. Comet C/1961~R1 has an abundance that slightly exceeds our reported values, but this is within error. We note that we exclude C/2016~R2  due to its peculiar water-poor nature. Similar to when considering the N$_2$/NH$_3$ ratio, the only other comet that significantly deviates from the experimental range of N$_2$/H$_2$O ratios is Hale-Bopp, which is depleted by over an order of magnitude. 
In the case of CO, only two comets (103P and 73P; \cite{le2015inventory} and references therein) have CO/H$_{2}$O values consistent with an origin from CO$_{2}$ photolysis/radiolysis. Most of the remainder, including 2013R1, Halley, 9P, 21P, Hale-Bopp, Hyakutake, C/2009 P1, and 2I/Borisov, probably require some other mechanism to explain their large CO reservoirs. There is, however, an intermediate set of comets, including 2P, 67P, 2001A2, and 2012F6, (\cite{le2015inventory} and references therein), which are within a factor of 5 of our predicted CO/H$_{2}$O abundances and considering the large uncertainties in the ice initial composition in the Solar Nebula, we cannot exclude a photolytic/radiolytic origin of CO in these comets.

A third molecule that may have a photolytic origin in comets is O$_2$
\citep{bieler2015abundant,mousis2016protosolar,bulak2022quantification}. Recently \citet{bulak2022quantification} investigated the production of O$_2$ in both H$_2$O and H$_2$O:CO$_2$ analogs, and reported final O$_2$/H$_2$O ratios of (0.9~$\pm$~0.2)\% and (1.6~$\pm$~0.4)\% from H$_2$O and H$_2$O:CO$_2$ ices, respectively, consistent with the Comet 67P O$_2$ abundance \citep{lauter2020gas}. The same experiments also produced substantial amounts of H$_2$O$_2$, which are not observed in 67P, challenging a photolytic explanation of O$_2$ and there are also other pieces of evidence suggesting that cometary O$_2$ may instead be inherited \citep{taquet2016primordial,taquet2017origin,taquet2018linking}. Additional experiments using more complex ice mixtures are likely required to resolve whether H$_2$O$_2$ is a necessary by product of O$_2$ formation, or whether more cometary molecules than N$_2$ may be consistent with a photolytic origin. The experiments presented here were not set up to monitor O$_2$ or H$_2$O$_2$ production, and can therefore not be used to either support or challenge this scenario.

In summary, both cometary N$_2$/NH$_3$ and N$_2$/H$_2$O ratios are consistent with a photolytic/radiolytic origin of N$_2$ when taking into account the possibility of N$_2$ outgassing and/or a range of 
initial NH$_3$ ice abundances. The 67P NH$_3$ abundance is about an order of magnitude lower than our fiducial experiments, and while this may simply reflect the end state of a photoprocessed ice (it is similar to our final ice composition), it may also be explained by a NH$_3$-poor initial ice. Based on an experiment of a NH$_3$-poorer ice, we find that N$_2$ production scales with initial NH$_3$ concentration. The distribution of cometary N$_2$ abundances may hence reflect comets forming from icy grains with a distribution of initial NH$_3$ abundances. 
By contrast only a small fraction of CO observations in comets are obviously compatible with an origin from CO$_2$ photolysis/radiolysis. For the specific case of 67P, the CO observations are marginally consistent with experiments: the CO/CO$_2$ ratios are similar to what is achieved experimentally in pure CO$_2$ ices, while the CO/H$_2$O ratios are a factor of 3 higher than the maximum seen in the experiments. We next turn to avenues to distinguish photolytic and entrapment origins of CO and N$_2$ in comets for the cases where a photolytic/radiolytic origin is a plausible explanation based on abundance ratios.

\subsection{Entrapment of Gas-Phase Molecules vs Photolysis Origins of Cometary CO and N$_2$}\label{subsec:entrapmentvsphoto}

Thus far, we have claimed that low levels of cometary hypervolatile origins can be explained by photochemical production, but another possibility is that the observed CO and N$_2$ are instead due to entrapment of interstellar or nebular CO and N$_2$ gas. Both water and CO$_2$ ices have been shown to efficiently entrap hypervolatile species \citep{bar1985trapping,simon2023entrapment}. As shown in the previous sub-section, photolysis in interstellar ice analogs can explain CO/H$_2$O abundances up to 1\%, and N$_2$/H$_2$O ratios up to 0.1\%. 
By contrast, entrapment of hypervolatiles from the gas-phase has been shown to achieve abundances with respect to water up to  $\sim$7~\% for CO and 1~\% for N$_2$ \citep{zhou2024competitive}. Entrapment at higher temperatures results in lower hypervolatile mixing ratios, and based on abundances alone, entrapment could explain almost all the observed cometary abundances. Only for the most CO and N$_2$-rich comets (including C/2009 P1, 2I/Borisov, and C/2016~R2) even entrapment does not suffice and these must have formed beyond the CO and N$_2$ snowlines \citep{price2021ice}. Using Occam's razor we may then wish to conclude that entrapment provides a sufficient explanation for all low- and mid-level hypervolatile cometary abundances.

There is, however, another cometary observation that could be used to distinguish between the two hypervolatile origin scenarios: isotopic ratios in hypervolatiles and possible parent molecules. If CO and N$_2$ originate from CO$_2$ and NH$_3$, respectively, we should expect similar $^{12}$C/$^{13}$C, $^{16}$O/$^{18}$O, and $^{14}$N/$^{15}$N ratios in the parent and daughter molecules. The isotopic ratios of $^{12}$C/$^{13}$C for CO and CO$_{2}$ in Comet 67P are indeed consistent with each other (84 $\pm$ 4 for $^{12}$CO$_{2}$/$^{13}$CO$_{2}$ and 86 $\pm$ 8 for $^{12}$CO/$^{13}$CO \citep{hassig2017isotopic,rubin2017evidence,nomura2022isotopic}). However, in the ISM CO$_{2}$ is expected to form from CO \citep{garrod2011formation}, and therefore similar isotopic ratios should be present even if CO is entrapped from the nebular gas, and hence the similar isotopic signatures of CO and CO$_{2}$ in comets are not very constraining.

By contrast, for N$_{2}$ we would expect different signatures if it originates from the gas-phase versus if it is formed in the ice through photolysis/radiolysis. Based on observations of Jupiter, which likely obtained its N reservoir from nebular N$_{2}$, nebular N$_{2}$ had a $^{14}$N/$^{15}$N isotopic ratio of $\sim$440, consistent with the bulk solar nebula value \citep{fletcher2014origin,nomura2022isotopic}. Cometary N$_2$ originating from nebular gas entrapment would be expected to have the same isotopic composition. Instead the $^{14}$N$_{2}$/$^{15}$N$_{2}$ ratio of Comet 67P has been reported as 130~$\pm$~30, which is consistent within uncertainties with the 67P $^{14}$NH$_{3}$/$^{15}$NH$_{3}$ ratio of  118~$\pm$~25 \citep{altwegg2019cometary,nomura2022isotopic}, disfavoring a nebular entrapment origin. Indeed, based on these values, \citet{nomura2022isotopic} proposed that N$_{2}$ does not exclusively originate from inheritance in 67P. Furthermore, it seems like all measured comet N-compounds have a similarly elevated $^{14}$N/$^{15}$N ratios; \citet{manfroid2009cn} investigated the C$^{14}$N/C$^{15}$N ratio in 18 comets including Halley and Hale-Bopp, and reported an average $^{14}$N/$^{15}$N ratio of 147.8 $\pm$ 5.7. The elevated values of N$_2$, NH$_3$, and CN are all consistent with each other, which supports a scenario where they all share a parent molecule. We suggest the parent molecule is likely NH$_3$, given its abundance in interstellar ices.  We note, however, that this isotopic data supports rather than proves our hypothesis. There are multiple elements in comets that show atypical isotopic ratios, including H, S, O and Xe \citep{nomura2022isotopic,calmonte2016sulphur,calmonte2017sulphur,marty2017xenon,altwegg2019cometary}, indicating that comets generally did not accrete material consistent with bulk Solar Nebula isotopic ratios. Nevertheless, based on this isotopic data, combined with the consistent N$_2$/NH$_3$ and N$_2$/H$_2$O abundances reported above, we propose that most cometary N$_{2}$, with the exception of hypervolatile dominated comets, could be due to NH$_{3}$ photolysis. Meanwhile for CO, entrapment is needed to explain CO/H$_{2}$O levels beyond 1\% and CO/CO$_{2}$ levels beyond 62\%, while photolysis and entrapment provide plausible explanations for lower levels of cometary CO.

Another observation that could potentially provide distinguishing evidence is the cometary desorption patterns of CO and N$_2$. In 67P both molecules appear to mostly desorb with water, while about a third is instead associated with CO$_2$ desorption. This is consistent with the nebular entrapment scenario, since both N$_2$ and CO may become entrapped in water and CO$_2$ ice matrices alike \citep{simon2023entrapment}. In our experiments, N$_2$ predominantly co-desorbs with water rather than with CO$_2$, and about half of the CO co-desorbs with water, and the remaining half co-desorbs with CO$_2$. Without further experiments exploring whether this depends on the precise matrix composition and thickness, we hesitate to use this as dispositive evidence for a photolytic origin, however.

Finally, it may at first seem strange to propose entrapment as an explanation for cometary CO and not for N$_2$, considering their similar sublimation temperatures and entrapment efficiencies. We speculate that cometary CO may originate from CO that co-deposits with water-ice formation in molecular clouds \citep{pontoppidan2003m,pontoppidan2006spatial}. By contrast to CO, which forms early in molecular clouds, N$_2$ formation chemistry is slow \citep{daranlot2012elemental,yamamoto2017introduction}, resulting in low N$_2$ gas abundances while water ice is forming on interstellar grains. This would imply that when interstellar grains enter a protoplanetary disk, some fraction of CO is protected inside the water-dominated ice layer of grains, while N$_2$ ice is only present in a water-poor ice and hence very vulnerable to sublimation. N$_2$ may therefore only become incorporated in the water-rich ice through the photolysis of NH$_3$ and related molecules.

\section{Conclusions} \label{sec:conclusions}
This study investigates the yields of CO and N$_{2}$ from photolysis of CO$_2$ and NH$_3$ in  different interstellar ice analogs of varying compositions inspired by interstellar ice observations (CO$_2$, NH$_3$, H$_2$O:CO$_2$, H$_2$O:NH$_3$, and H$_2$O:CO$_2$:NH$_3$) and temperatures (10--100~K), and exposed to two different dissociation sources (UV and electrons). We obtained the yields with respect to parent molecules as well as to water after an energy dose commensurate to the expected exposure for icy grains in molecular clouds. Our main findings are:
\begin{enumerate}
    \item Qualitatively, CO formed in all experiments containing CO$_{2}$, and N$_{2}$ in all containing NH$_3$. Quantitatively, the undiluted experiments resulted in final mixing ratios such that the CO/CO$_{2(f)}$ ratio is 51--62~\%, and N$_{2}$/NH$_{3(f)} \sim$15-21~\%. Meanwhile, the water-rich experiments yielded a range of final mixing ratios such that the CO/CO$_{2(f)}$ ratio is 3--10~\%, N$_{2}$/NH$_{3(f)} \sim$4--19~\%, CO/H$_{2}$O$ \sim$0.4--0.9~\%, and N$_{2}$/H$_{2}$O $\sim$0.03--0.7~\%.
    \item In both UV and electron radiation experiments, the CO and N$_2$ yields were higher in the undiluted ices compared to water-rich binary and ternary ices. In the former, up to 25\% of the initial CO$_{2}$ was converted into CO, and up to 18\% of the initial NH$_{3}$ into N$_{2}$ when considering two NH$_{3}$ molecules are required to form one N$_{2}$ molecule. Meanwhile, in water-rich ices, up to 8\% of the initial CO$_{2}$ was converted into CO, and up to 4\% of the initial NH$_{3}$ into N$_{2}$ taking into account that two NH$_{3}$ are needed to form one N$_{2}$. Interestingly the CO yield increases when adding NH$_3$ to the ice, perhaps due to a softening effect that reduces the cage effect during CO$_2$ photolysis, while the N$_2$ yield decreases in ices with CO$_2$, likely due to an increased formation of other N-containing molecules than N$_2$.
    \item In ternary H$_2$O:CO$_2$:NH$_3$ $\sim$100:20:5 ices, the CO and N$_{2}$ yields increase by a factor of a few with temperature, except for an apparent local minimum at 30~K. 
    \item The CO yields are similar between the UV and electron bombardment experiments. However, the N$_2$ yields were substantially higher in the electron bombardment experiments compared to the UV photolysis experiments. Though in both cases the experiments appear to have reached steady state and should hence be comparable, we cannot rule out that this is in part explained by a higher energy dose in the electron bombardment experiments.
    \item We do not produce sufficient CO from our experiments to suggest that cometary CO is primarily produced from photoprocessing; our CO yields are consistent with a few comets that have very low CO abundances, including 103P and 73P. For most comets with a higher CO abundance (including 67P), entrapment remains to the most plausible origin.
    \item By contrast sufficient N$_2$ is produced in the photolysis and radiolysis experiments to explain most of the observed cometary N$_2$. This conclusion is corroborated by the isotopic similarity of N$_2$ and NH$_3$ in 67P. The measured N$_2$ abundances in comets should be used cautiously when deducing the thermal history of the comet-forming material.
\end{enumerate}
\section{Acknowledgements} \label{sec:ack}
A.M. thanks Dr. S. Narayanan for technical training with SPACETIGER. A.M. acknowledges the support of the Natural Sciences and Engineering Research Council of Canada (NSERC), [funding reference number 587448]. Cette recherche a été financée par le Conseil de recherches en sciences naturelles et en génie du Canada (CRSNG), [numéro de référence 587448]. This work is also supported from a Simons Foundation grant (grant No. 686302 - K.I.Ö.).
\bibliography{CO_and_N2.bib}{}
\bibliographystyle{aasjournalv7}

\appendix{}\label{sec:appendix}
\section{$^{12}$CO$_{2}$ Undiluted ice UV Irradiation Results}\label{app:12CO}
\begin{figure*}[ht]
\centering
  \includegraphics[width=12cm]{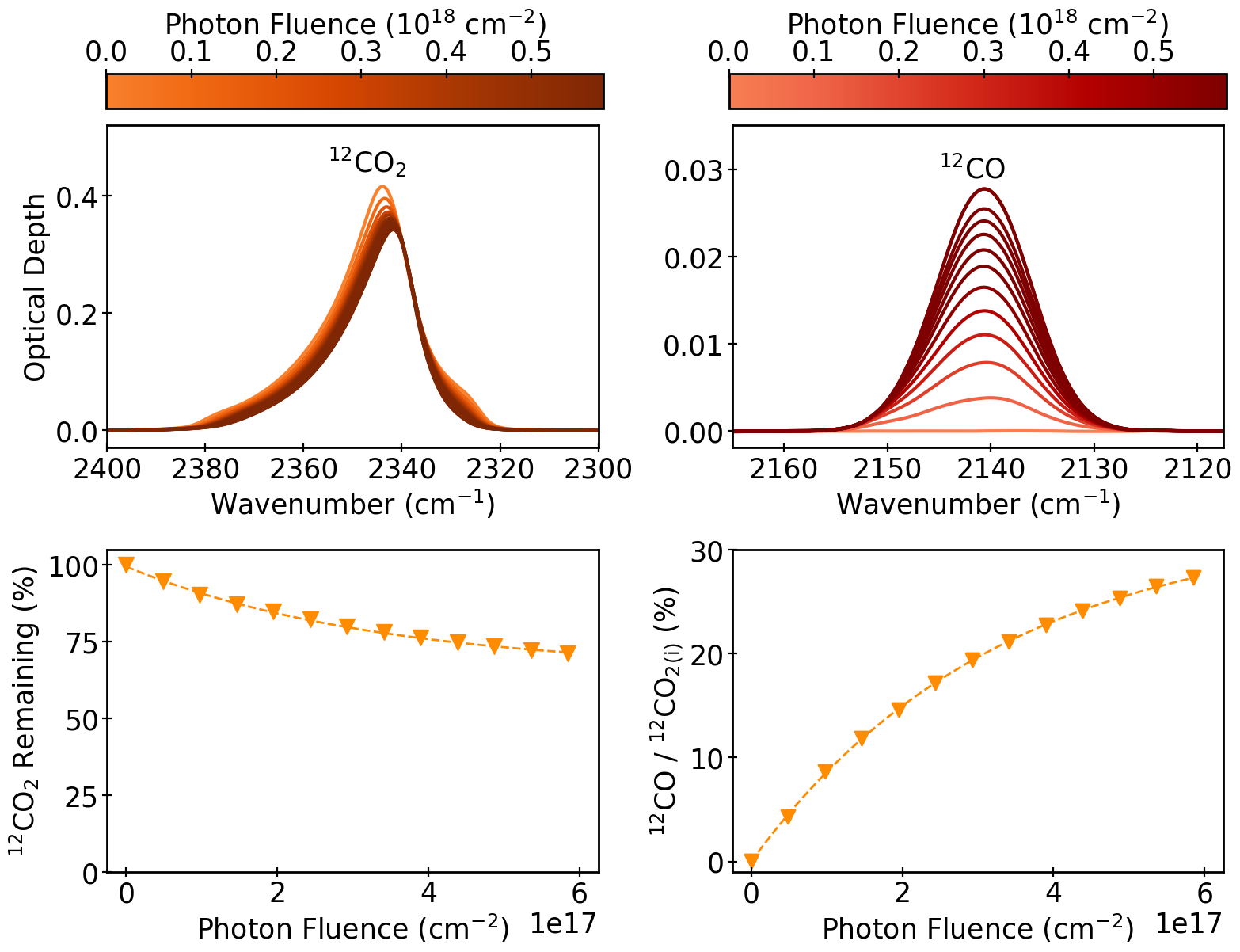}
  \caption{Top: The change in the spectral features throughout irradiation of a $^{12}$CO$_{2}$ undiluted ice. Bottom: The $^{12}$CO$_{2}$ destruction (left) and CO formation (right) throughout irradiation of a $^{12}$CO$_{2}$ undiluted ice.}
  \label{fgr:12CO2Temporal}
\end{figure*}

Figure \ref{fgr:12CO2Temporal} shows the results from a $^{12}$CO$_{2}$ Undiluted ice UV experiment to confirm that a change in isotopologue does not have a significant impact on our results. The initial ice was 82 ML thick. We present the spectral features we are integrating, as well as the CO$_2$ destruction and CO formation curves.
The $^{12}$CO$_{2}$ destruction cross section is (3~$\pm$~1)$\times$10$^{-18}$~cm$^{-2}$~s$^{-1}$ and the $^{12}$CO formation cross section is also (3~$\pm$~1)$\times$10$^{-18}$~cm$^{-2}$~s$^{-1}$ Following the VUV destruction of $^{12}$CO$_{2}$, we produced 27~$\pm$~8 \% $^{12}$CO$_{2}$ with respect to initial $^{12}$CO$_{2}$. This is within the error of what was seen for the undiluted CO$_{2}$ ices in Section~\ref{subsec:Primary}.
\clearpage

\section{IR Spectra}\label{app:IR_Spectra}

\begin{figure*}[ht]
\centering
  \includegraphics[width=18cm]{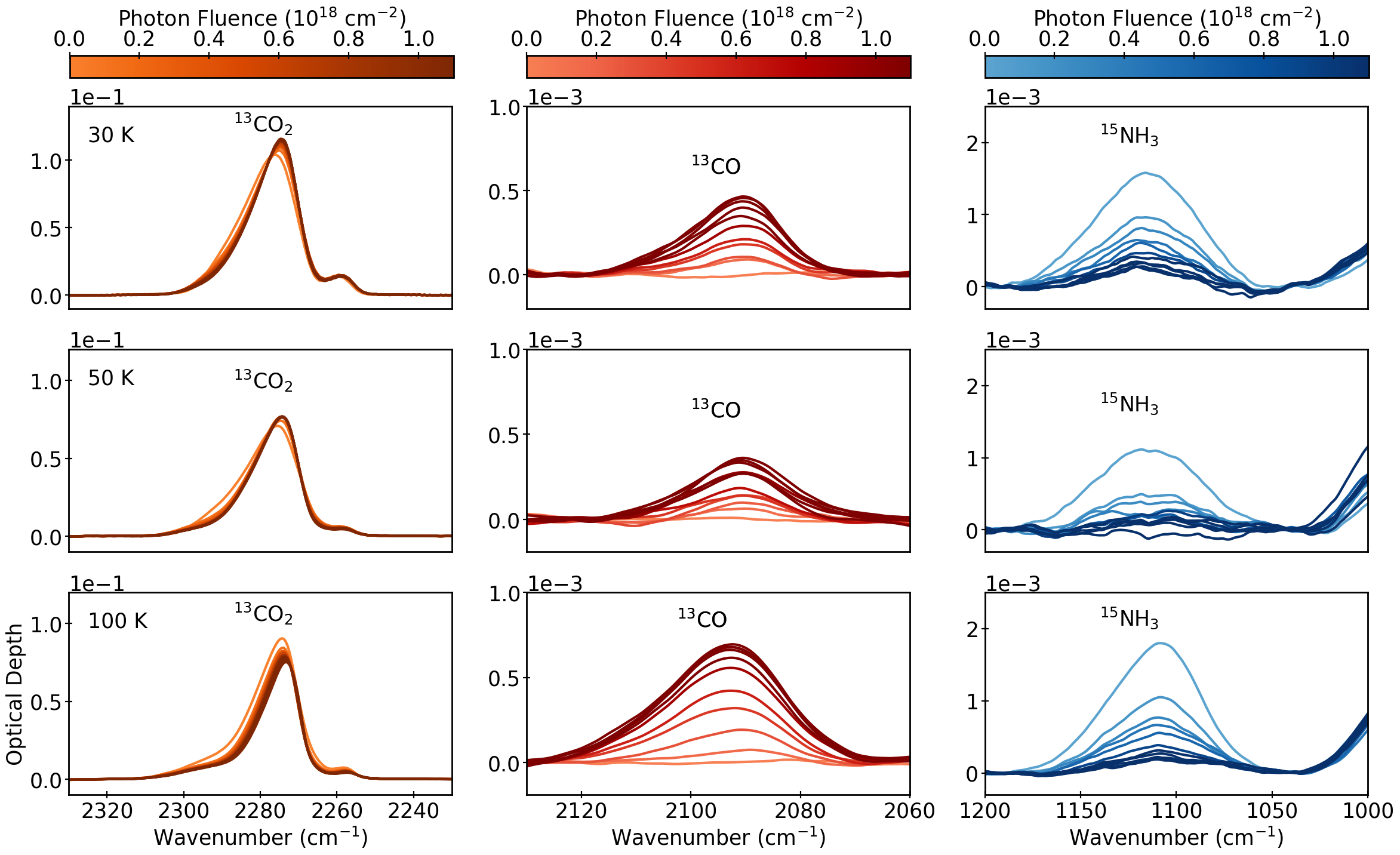}
  \caption{The infrared spectral bands  of CO$_2$ (left), CO (middle) and NH$_3$ (right) during photolysis of  ternary ices at 30 (top), 50 (middle) and 100 K (bottom).  The baselines used here are the same as those described in Section~\ref{sec:IR}.}
  \label{fgr:Change_In_Peak_Temp}
\end{figure*}

\begin{figure*}[ht]
\centering
  \includegraphics[width=18cm]{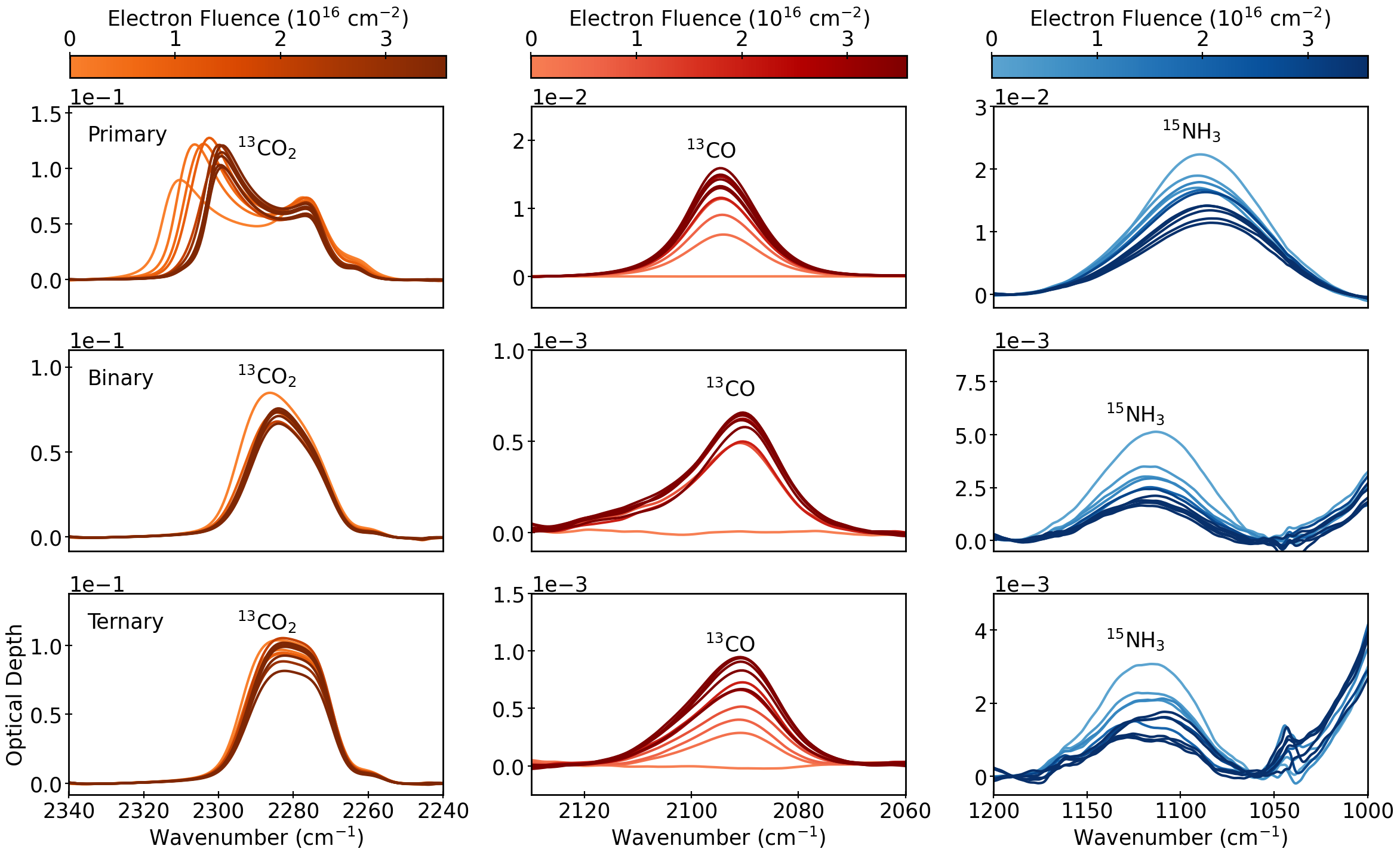}
  \caption{The infrared spectral bands  of CO$_2$ (left), CO (middle) and NH$_3$ (right) during electron bombardment of  primary ices (top), binary ices (middle) and ternary ices (bottom).  The baselines used here are the same as those described in Section~\ref{sec:IR}.}
  \label{fgr:Change_In_Peak_Elec}
\end{figure*}

Figures \ref{fgr:Change_In_Peak_Temp} and \ref{fgr:Change_In_Peak_Elec} show the spectral changes of CO$_2$, NH$_3$ and CO IR bands during UV exposure at different temperatures and electron exposure at 10 K, respectively. We note that due to artifacts in the infrared spectra  during electron bombardment, the NH$_{3}$ peak is difficult to integrate as it sometimes overlaps with the artifact. 

\clearpage

\section{Kinetic Fit Parameters}\label{app:FitParameters}

Table \ref{tbl:fits} shows the cross sections and steady state column densities for CO$_{2}$, NH$_{3}$, and CO for all experiments.

\begin{deluxetable}{lc||cc|cc|cc}[ht]					
\tabletypesize{\scriptsize}					
\tablewidth{0pt}					
%\tablenum{1}					
\tablecaption{The kinetic model fit parameters used for this study \label{tbl:fits}}					
\tablehead{					
\colhead{} &\colhead{} &  \multicolumn{2}{c}{CO$_{2}$} &  \multicolumn{2}{c}{NH$_{3}$}   & \multicolumn{2}{c}{CO}\\					
\colhead{Molecules} & \colhead{Irr} & \colhead{$\sigma_d$ (10$^{-18}$)} & \colhead{$N_{ss}$} & \colhead{$\sigma_d$ (10$^{-18}$)} & \colhead{$N_{ss}$}  & \colhead{$\sigma_f$ (10$^{-18}$)} & \colhead{$N_{ss}$} \\					
\colhead{} & \colhead{Temp (K)}  & \colhead{(cm$^{-2}$s$^{-1}$)} & \colhead{(ML)} & \colhead{(cm$^{-2}$s$^{-1}$)} & \colhead{(ML)}  & \colhead{(cm$^{-2}$s$^{-1}$)} & \colhead{(ML)}					
}					
\startdata							
SPACECAT \\					
\hline								
CO$_{2}$	&	N/A	\\								
CO$_{2}$	&	10	&	2.5	$\pm$	0.4	&	10	$\pm$	2	&				&				&	4	$\pm$	1	&	18	$\pm$	5	\\
\hline											
NH$_{3}$	&	N/A	\\								
NH$_{3}$	&	10	&				&				&	18	$\pm$	7	&	2	$\pm$	1	\\				
\hline											
H$_{2}$O:CO$_{2}$	&	N/A	\\%doublecheckthis	&									
H$_{2}$O:CO$_{2}$	&	10	&	1.6	$\pm$	0.3	&	16	$\pm$	3	&				&				&	8	$\pm$	3	&	1.2	$\pm$	0.3	\\
\hline											
H$_{2}$O:NH$_{3}$	&	N/A	\\%doublecheckthis	&									
H$_{2}$O:NH$_{3}$	&	10	&				&				&	5	$\pm$	1	&	4	$\pm$	2	\\	\hline												
H$_{2}$O:CO$_{2}$:NH$_{3}$	&	N/A	\\	
H$_{2}$O:CO$_{2}$:NH$_{3}$$^b$	&	10	&	1.1	$\pm$	0.7	&	20	$\pm$	4	&	8	$\pm$	3	&	0.3	$\pm$	0.3	&	3	$\pm$	1	&	3.0	$\pm$	0.8	\\
H$_{2}$O:CO$_{2}$:NH$_{3}$$^b$	&	10	&	1.0	$\pm$	0.2	&	17	$\pm$	4	&	6	$\pm$	2	&	0.4	$\pm$	0.2	&	1.8	$\pm$	0.7	&	1.8	$\pm$	0.5	\\
H$_{2}$O:CO$_{2}$:NH$_{3}$$^a$	&	10	&	0.9	$\pm$	0.5	&	14	$\pm$	3	&	5	$\pm$	1	&	0.8	$\pm$	0.5	&	1.9	$\pm$	0.7	&	2.1	$\pm$	0.6	\\
H$_{2}$O:CO$_{2}$:NH$_{3}$	&	30	&	3.0	$\pm$	0.4	&	15	$\pm$	3	&	7	$\pm$	2	&	0.9	$\pm$	0.5	&	1.4	$\pm$	0.6	&	1.5	$\pm$	0.4	\\%2025/04/08
H$_{2}$O:CO$_{2}$:NH$_{3}$	&	50	&	3.2	$\pm$	0.6	&	11	$\pm$	2	&	11	$\pm$	4	&	0.9	$\pm$	0.5	&	2.2	$\pm$	0.9	&	2.4	$\pm$	0.6	\\
H$_{2}$O:CO$_{2}$:NH$_{3}$	&	100	&	6.0	$\pm$	1.2	&	10	$\pm$	2	&	6	$\pm$	2	&	0.7	$\pm$	0.4	&	3	$\pm$	1	&	2.7	$\pm$	0.7	\\
\hline										
SPACETIGER	&	e$^{-}$	Dose($\times$10$^{16}$)	\\					
\hline											
CO$_{2}$	&	10	&	112	$\pm$	52	&	$\sim$~12	&				&				&	360	$\pm$	150	&	$\sim$~24	\\
\hline											
NH$_{3}$	&	10	&				&				&	31	$\pm$	21	&	$\sim$~21	\\								
\hline											
H$_{2}$O:CO$_{2}$	&	10	&	340	$\pm$	170	&	$\sim$~7	&				&				&	320	$\pm$	130	&	$\sim$~0.9	\\
\hline											
H$_{2}$O:NH$_{3}$	&	10	&				&				&	380	$\pm$	110	&	$\sim$~4	\\								
\hline											
H$_{2}$O:CO$_{2}$:NH$_{3}$	&	10	&	320	$\pm$	330	&	$\sim$~11	&	280	$\pm$	80	&	$\sim$~2	&	190	$\pm$	82	&	$\sim$~1.8	\\
\enddata					
\tablecomments{Our error propagation is assuming symmetric errors, which in some cases result in unphysical error bars.\\  $^c$The obtained cross sections for the electron bombardment experiments have different units of electron$^{-1}$$\times$ cm$^{-2}$$\times$ s$^{-1}$. $^a$Fiducial Experiment; $^b$Repeat Experiment.}	
\end{deluxetable}

The uncertainties of the destruction cross sections in Equation~\ref{eqn:decay} and the formation cross sections in Equation~\ref{eqn:production} from the fit to the data are combined with the uncertainties from experimental variability and the lamp flux calibration to yield the total parameter uncertainties:

\begin{equation}
   \sigma_{\sigma_{x}} = \sqrt{(\sigma_{Kineticfit})^2+(\sigma_{exp})^2+(\sigma_{Lamp})^2}
\end{equation}

The experimental variability in the cross section from repeat experiments are $\sim$10\% for CO$_2$, $\sim$24\% for NH$_3$, and 38\% for CO. Adding the uncertainties in quadrature, we obtain a final uncertainty of  $\sim$11\% for CO$_2$, $\sim$25\% for NH$_3$, and $\sim$38\% for CO. 

The uncertainties in the fitted steady state column densities can be calculated as:
\begin{equation}
   \sigma_{N\mathrm{_{ss}}(x)} = \sqrt{(\sigma_{Kineticfit})^2+(\sigma_{A})^2+(\sigma_{exp})^2}
\end{equation}
in which $\sigma_{Kineticfit}$ is the model fit uncertainty, which is typically $<$1\%, and the uncertainty from experimental variability $\sigma_{exp}$ is 17\% for CO, 4\% for CO$_2$, and 52\% for NH$_3$ based on repeat experiments. The band strength uncertainty for all species is assumed to be 20\%. For most experiments we obtain an overall uncertainty of $\sim$21\% for CO$_2$, $\sim$55\% for NH$_3$, and $\sim$27\% for CO.

\section{TPD Curves}\label{app:TPDs}

\begin{figure*}[ht]
\centering
   \includegraphics[width=18cm]{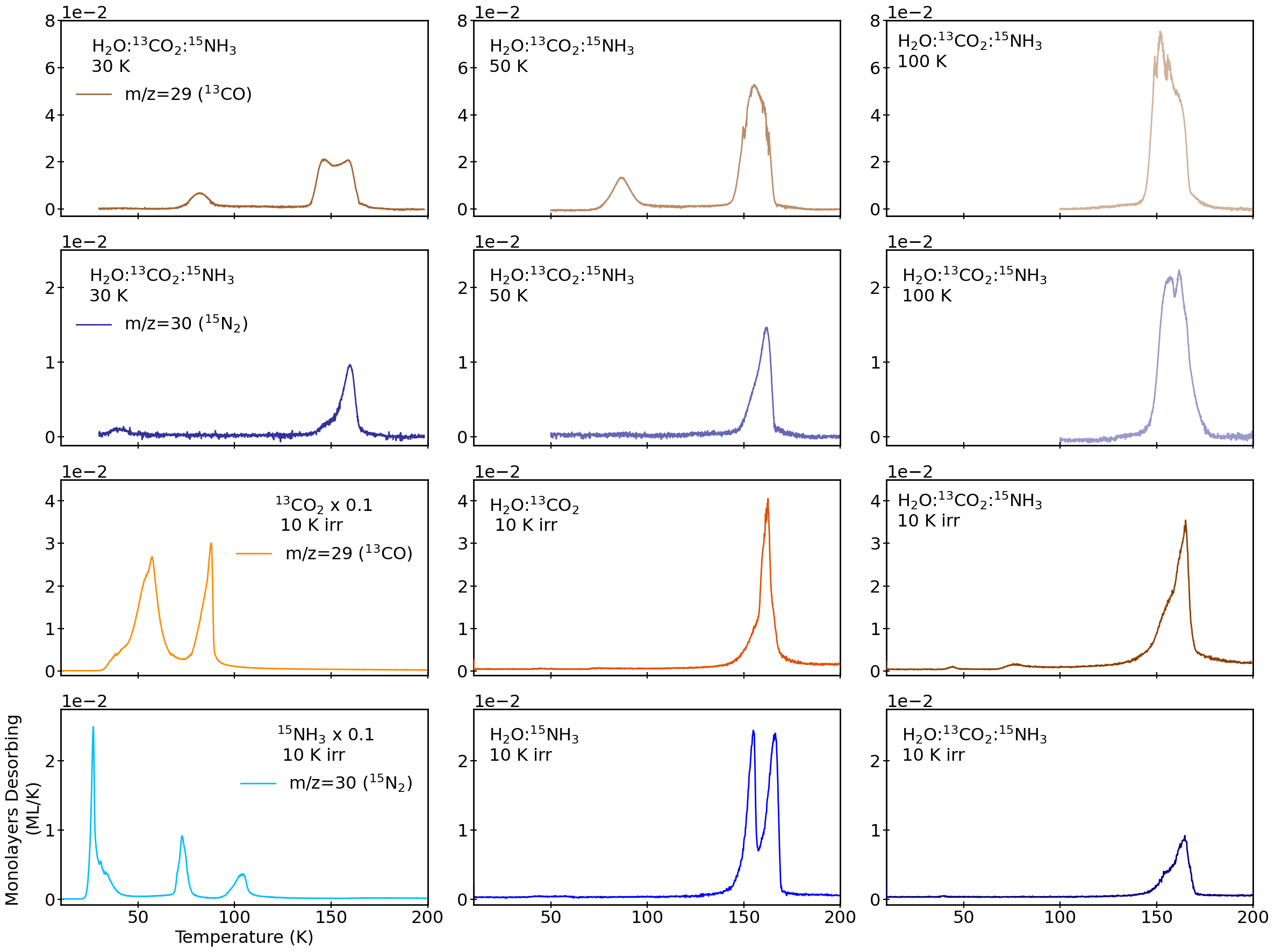}
  \caption{Top two rows: The Temperature Programmed Desorption (TPD) of $^{15}$N$_{2}$ (m/z=30) and $^{13}$CO (m/z=29). From left to right: 30~K irradiation; 50~K irradiation; 100~K irradiation. Bottom two rows: The Temperature Programmed Desorption (TPD) of $^{13}$CO (m/z=29; top) and of $^{15}$N$_{2}$ (m/z=30; bottom) following electron bombardment. From left to right: Primary, Binary, Ternary. Note that for the two primary electron bombardment experiments, y-axis is scaled by a factor of 0.1.}
  \label{fgr:TPD_Temp_and_Electrons}
\end{figure*}

\begin{figure*}[htbp]
\centering
  \includegraphics[width=18cm]{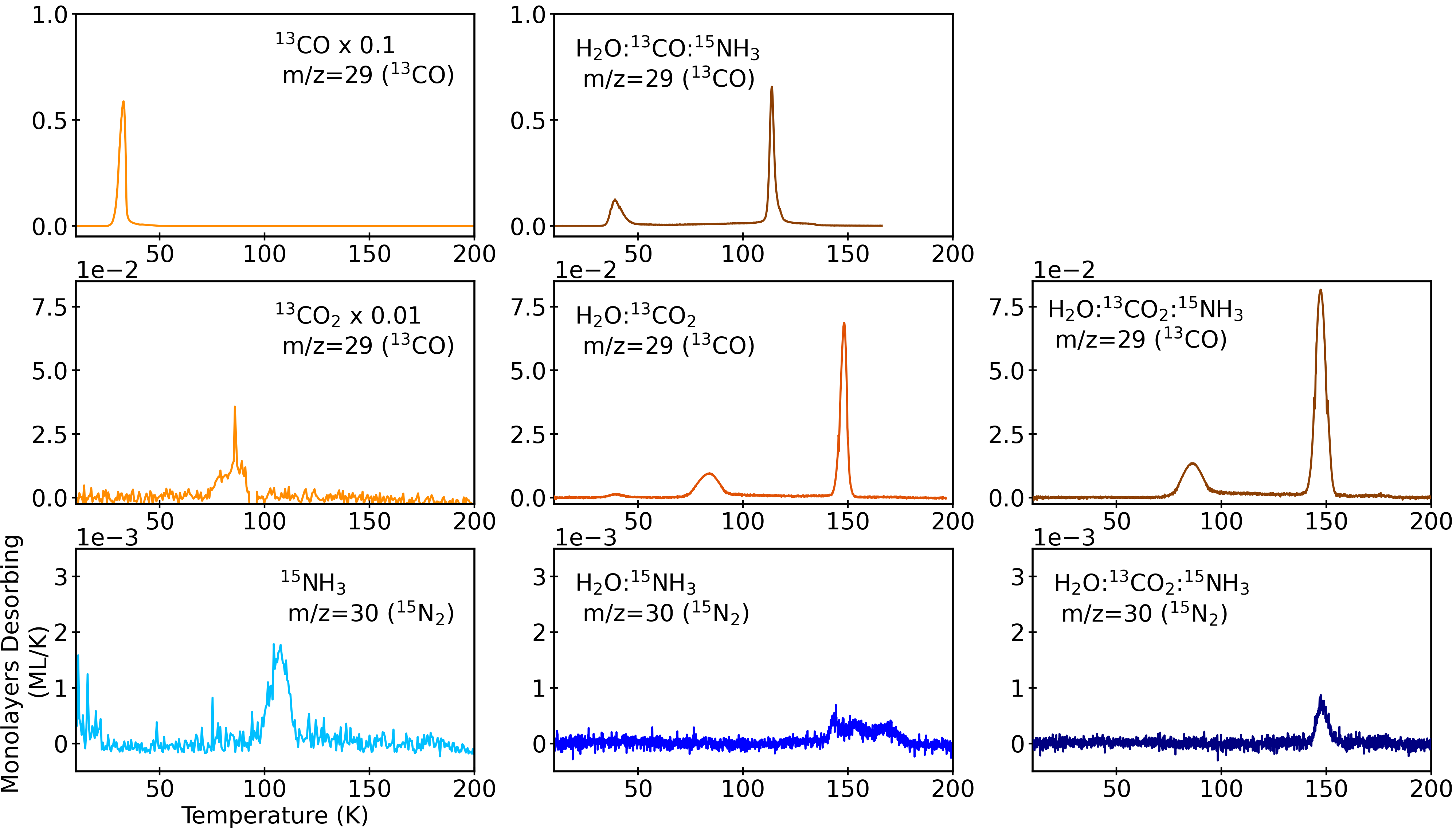}
  \caption{Top row: TPD of $^{13}$CO (m/z=29) used to calculate $k_{\mathrm{CO}}$ for UV experiments (left) and electron experiments (right). Bottom two rows: the Temperature Programmed Desorption (TPD) of $^{13}$CO (m/z=29; middle) and of $^{15}$N$_{2}$ (m/z=30; bottom) without any irradiation. From left to right: Primary, Binary, Ternary. Some of the TPD curves have been scaled for readability, which is listed on the individual panel.}
  \label{fgr:TPDsBlank}
\end{figure*}

Figure~\ref{fgr:TPD_Temp_and_Electrons} shows the N$_2$ and CO TPD curves for photolysis experiments at 30--100~K, and the three radiolysis experiments.  Figure~\ref{fgr:TPDsBlank} shows TPDs of experiments that were not exposed to any radiation, which were used to check the purity of the samples as well as to calculate $k_{\mathrm{^{13}CO}}$.  From the middle three panels of Figure~\ref{fgr:TPDsBlank}, we can see the $^{13}$CO originating from dissociation of $^{13}$CO$_{2}$ in the QMS. The bottom row shows that in the case of NH$_3$, the N$_2$ signal is an order of magnitude smaller than the m/z~=~30 signal from the irradiated $^{15}$NH$_{3}$-containing ices, and there is hence minimal $^{15}$N$_2$ contamination. The top row of Figure~\ref{fgr:TPDsBlank} shows the m/z~=~29 signal in a pure $^{13}$CO TPD used to determine the $k_{\mathrm{^{13}CO}}$ value by simultaneously measuring the ice thickness to 23 ML with the IR.

\newpage

\clearpage
\section{Repeat Experiment }\label{app:Repeat}

\begin{figure*}[ht]
\centering
  \includegraphics[width=18cm]{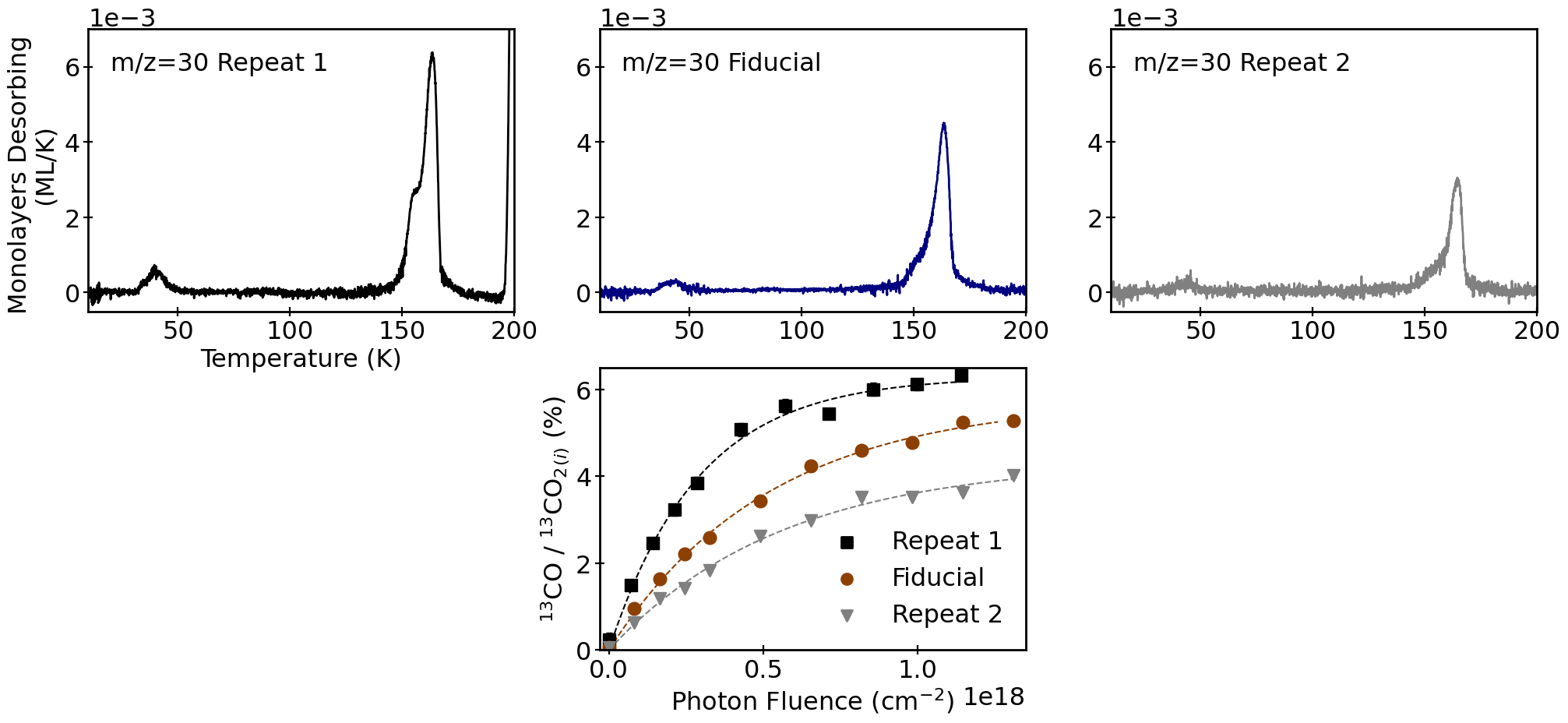}
  \caption{Top: The three $^{15}$N$_2$ TPD curves from the three repeat experiments. Bottom: The three CO formation curves from the three repeat experiments. }
  \label{fgr:Repeats_Fiducial_Exp}
\end{figure*}

Figure \ref{fgr:Repeats_Fiducial_Exp} shows  the N$_2$ TPDs and $^{13}$CO/$^{13}$CO$_{2,(\textit{i})}$  column density ratios (throughout irradiation) of three repeats of the fiducial 10~K ternary experiment. The resulting N$_2$ yields with respect to the inital NH$_3$ for the fiducial experiment, repeat 1, and repeat 2, are 1.0 \%, 1.1 \%, and 0.8 \%, respectively, resulting in a standard deviation of $\sim$20\%. The CO yields with respect to initial CO$_2$ for the fiducial experiment, repeat 1, and repeat 2, are 5.3 \%, 6.0 \%, and 4.0 \%, respectively, also resulting in a standard deviation of $\sim$20 \%

\clearpage

\section{Irradiation of a NH$_3$-Rich 7:1:1 H$_{2}$O:CO$_{2}$:$^{15}$NH$_{3}$ Ice and a 88:12:1 H$_{2}$O:CO$_{2}$:$^{15}$NH$_{3}$ Ice}\label{app:TwiceAmmonia}

\begin{figure*}[ht]
\centering
  \includegraphics[width=18cm]{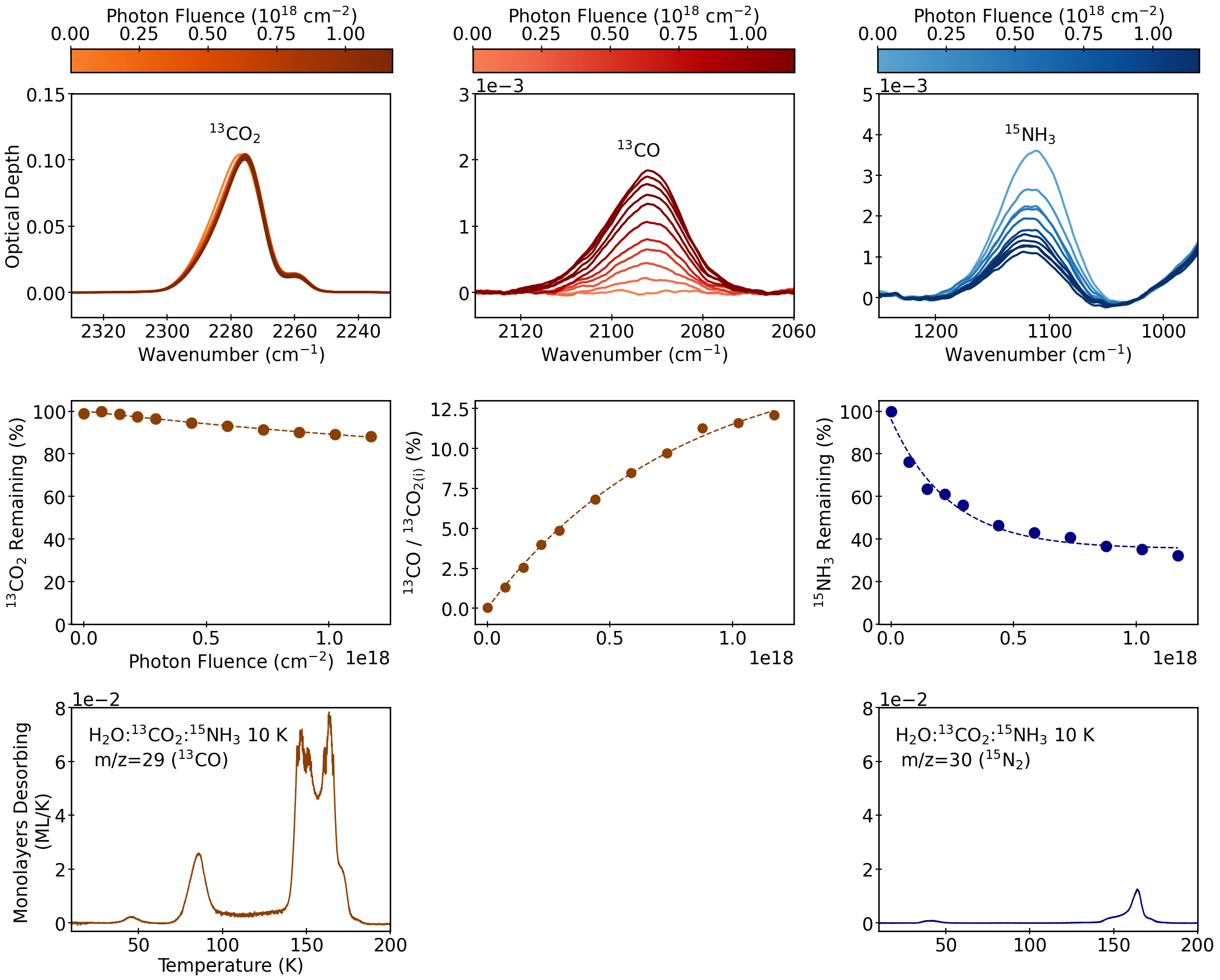}
  \caption{Top panels: The change in the IR features throughout irradiation of a NH$_3$-rich  ternary ice. Middle panels: CO$_{2}$ destruction, CO formation, and NH$_{3}$ destruction curves. Lower panels: TPD curves used to measure the resulting N$_2$ abundance.}
  \label{fgr:Results_2x_Ammonia}
\end{figure*}

We irradiated an ice similar to the fiducial case except that we increase the ammonia concentration by a factor of two. The initial ice was 99:18:14~ML H$_{2}$O:CO$_{2}$:NH$_{3}$ thick. Figure \ref{fgr:Results_2x_Ammonia} shows the spectral time series, the NH$_3$ and CO$_2$ destruction curves, the CO formation curve, and the m/z=29 and 30 TPDs. The experiment yielded (2.2~$\pm$~0.6)~ML CO formation, equivalent to (12~$\pm$~3)~\% with respect to the initial CO$_2$.

\begin{figure*}[ht]
\centering
  \includegraphics[width=18cm]{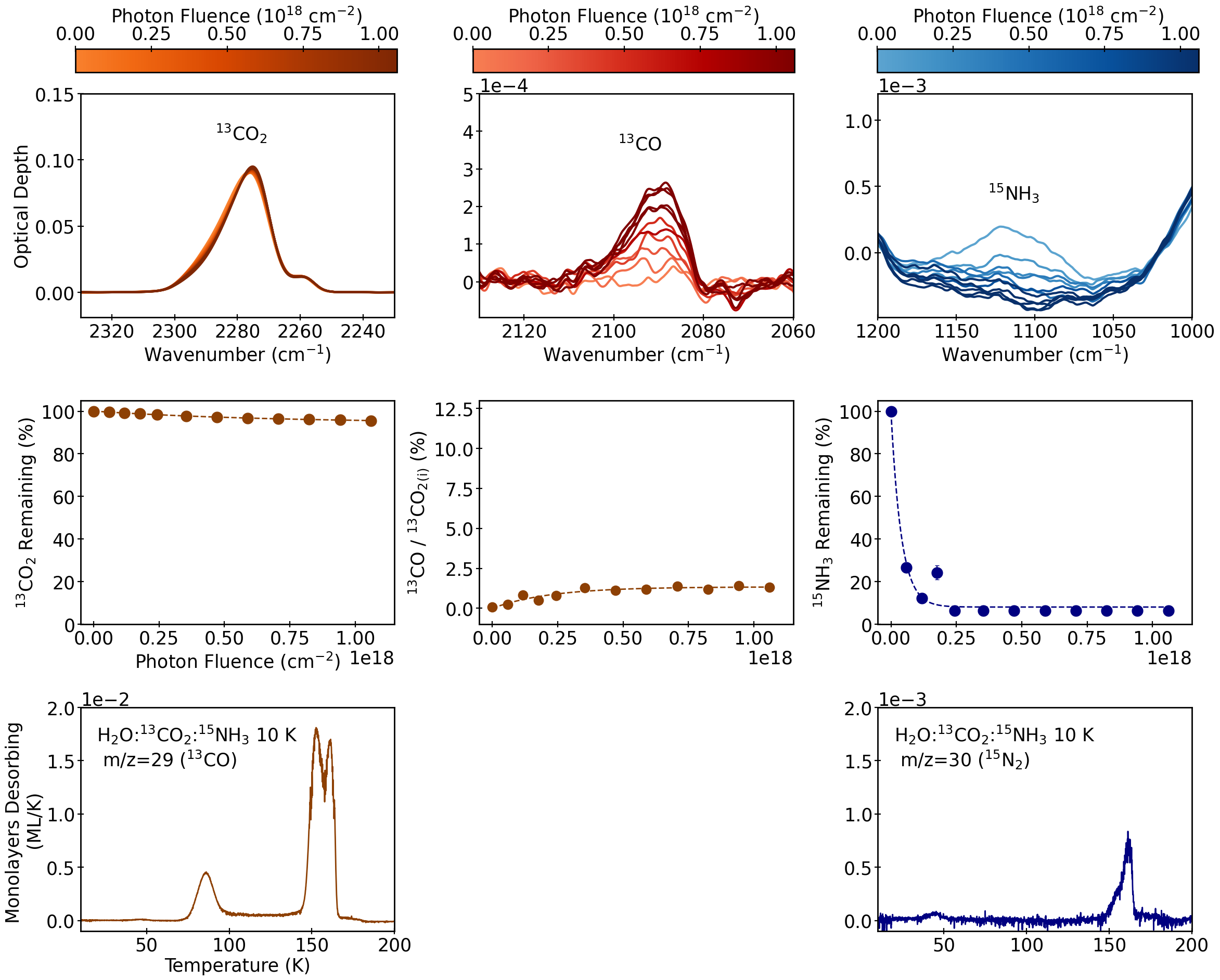}
  \caption{Top panels: The change in the IR features throughout irradiation of a NH$_3$-poor  ternary ice. Middle panels: CO$_{2}$ destruction, CO formation, and NH$_{3}$ destruction curves. Lower panels: TPD curves used to measure the resulting N$_2$ abundance.}
  \label{fgr:Results_1_2x_Ammonia}
\end{figure*}

\newpage
We also irradiated an additional ice similar to the fiducial case except that this time decrease the ammonia concentration by a factor of two. The initial ice was 114:16:1~ML H$_{2}$O:CO$_{2}$:NH$_{3}$ thick. Figure \ref{fgr:Results_1_2x_Ammonia} shows the spectral time series, the NH$_3$ and CO$_2$ destruction curves, the CO formation curve, and the m/z=29 and 30 TPDs. The experiment yielded (0.01~$\pm$~0.003)~ML N$_2$ formation, equivalent to (0.009~$\pm$~0.004)~\% with respect to water.

\clearpage
\section{Destruction and Formation Curves From Electron Bombardment Experiments}\label{app:SpaceTigerTemporalBehaviour}

\begin{figure}[ht]
\centering
  \includegraphics[width=8cm]{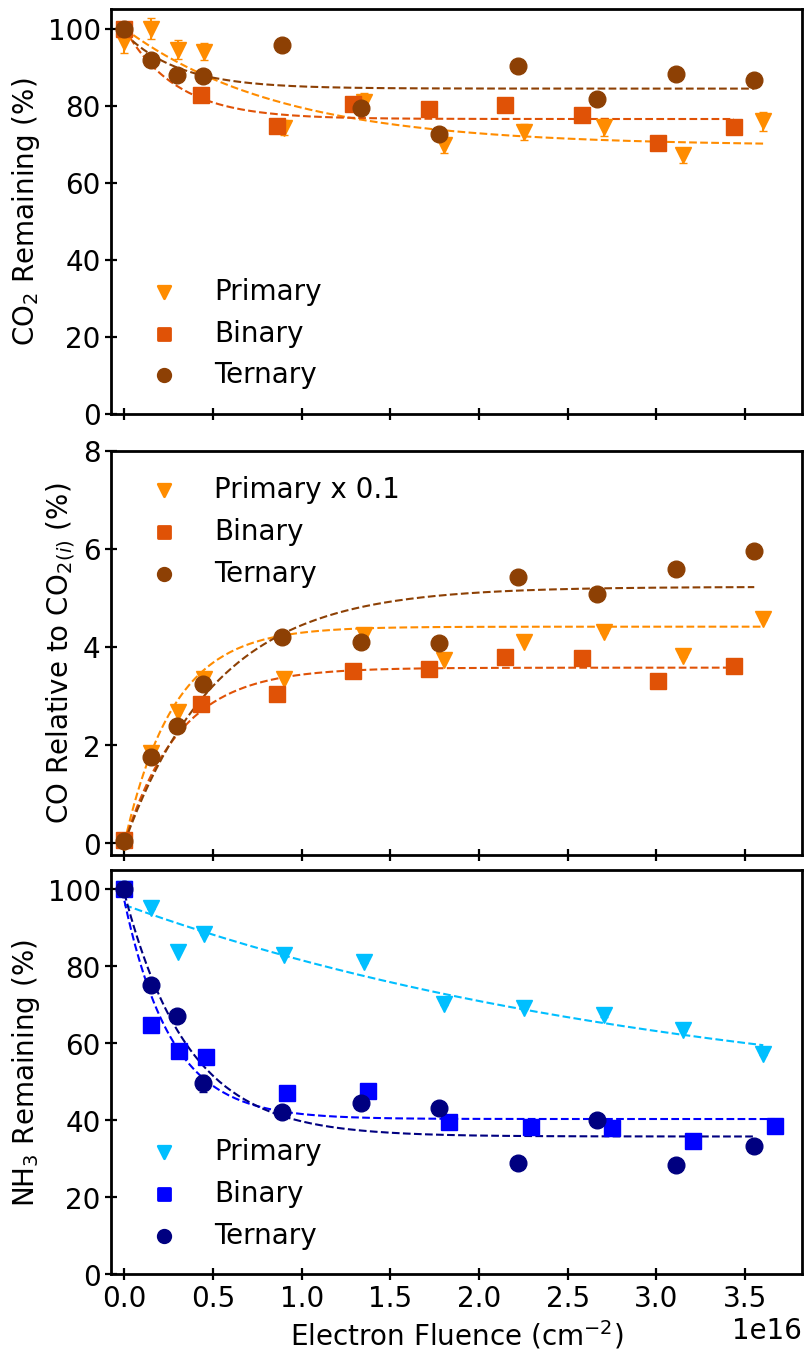}
  \caption{The temporal behavior of CO$_{2}$ (top panel) CO (middle panel) and NH$_{3}$ (bottom panel) following electron bombardment of ternary ices at various compositions. Note that in the CO trace panel (middle), the primary composition has been multiplied by a factor of 0.1 to clearly show the binary and ternary mixtures.}
  \label{fgr:ST_Temporal_Behaviour}
\end{figure}

Figure~\ref{fgr:ST_Temporal_Behaviour} shows the formation curves of CO as well as the destruction curves of CO$_{2}$ and NH$_{3}$ during electron bombardment.

\end{document}